\documentclass{aa}  
\usepackage{graphicx}
\usepackage{subfigure}
\usepackage{txfonts}
\usepackage{verbatim}
\usepackage{soul}
\usepackage{float}
\usepackage{pifont}
\usepackage{tabularx,adjustbox,booktabs}
\begin{document}

\title{Asteroseismological analysis of the ultra-massive ZZ Ceti stars BPM~37093, GD~518, and SDSS~J0840+5222}

\author{Alejandro H. C\'orsico\inst{1,2}, Francisco C. De Ger\'onimo\inst{1,2}, Mar\'ia E. Camisassa\inst{1,2}, Leandro G. Althaus\inst{1,2}}
\institute{$^1$Grupo de Evoluci\'on Estelar y Pulsaciones. Facultad de 
           Ciencias Astron\'omicas y Geof\'{\i}sicas, 
           Universidad Nacional de La Plata, 
           Paseo del Bosque s/n, 
           (1900) La Plata, 
           Argentina\\
       $^2$Instituto de Astrof\'{\i}sica La Plata, 
           IALP (CCT La Plata), 
           CONICET-UNLP\\
\email{acorsico@fcaglp.unlp.edu.ar}}
\date{Received ; accepted }

\abstract{Ultra-massive ($\gtrsim 1 M_{\sun}$) hydrogen-rich (DA)
  white dwarfs are expected  to  have  a  substantial portion of
  their cores in a crystalline state at the effective  temperatures
  characterizing the ZZ Ceti instability strip ($T_{\rm eff} \sim
  12\,500$ K), as a result of Coulomb interactions in very dense
  plasmas. Asteroseismological analyses of these white dwarfs can
  provide valuable information related to the crystallization process,
  the core chemical composition and the evolutionary origin of these
  stars.}{We present a thorough asteroseismological analysis of the
  ultra-massive ZZ Ceti star BPM~37093, which exhibits a rich period
  spectrum, on the basis of a complete set of fully evolutionary
  models that represent ultra-massive oxygen/neon(ONe)-core DA white
  dwarf stars harbouring a range of hydrogen (H) envelope
  thicknesses. We also carry out preliminary asteroseismological
  inferences on two other ultra-massive ZZ Ceti stars that exhibit
  fewer periods, GD~518, and SDSS~J0840+5222.}{We considered $g$-mode
  adiabatic pulsation periods for ultra-massive ONe-core DA white
  dwarf models with stellar masses in the range $1.10 \lesssim
  M_{\star}/ M_{\sun} \lesssim 1.29$,  effective  temperatures  in
  the  range  $10\,000 \lesssim T_{\rm eff} \lesssim 15\,000$ K,  and
  H  envelope  thicknesses  in  the  interval $-10 \lesssim
  \log(M_{\rm H}/M_{\star}) \lesssim -6$. We explore the effects of
  employing different H-envelope thicknesses on the mode-trapping
  properties of our ultra-massive ONe-core DA white dwarf models, and
  perform period-to-period fits to ultra-massive ZZ Ceti stars with
  the aim of finding an asteroseismological model for each target
  star.}{We found that the trapping cycle and trapping amplitude are
  larger for thinner H envelopes, and that the asymptotic period
  spacing is longer for thinner H envelopes. We found a mean period
  spacing of $\Delta \Pi \sim 17$ s in the data of BPM~37093, which is
  likely to be associated to $\ell= 2$ modes. However, we are not able
  to put constraints on the stellar mass of BPM~37093 using this mean
  period spacing due to the simultaneous sensitivity of $\Delta \Pi$
  with $M_{\star}$, $T_{\rm eff}$, and $M_{\rm H}$, an intrinsic
  property of DAV stars. We found asteroseismological models for the
  three objects under analysis, two of them (BPM~37093 and GD~518)
  characterized by canonical (thick) H envelopes, and the third one
  (SDSS~J0840+5222) with a thinner H envelope. The effective
  temperature and stellar mass of these models are in agreement with
  the spectroscopic determinations. The percentage of crystallized
  mass of these asteroseismological models is 92 \%, 97 \%, and 81 \%
  for BPM~37093, GD~518, and SDSS~J0840+5222, respectively. We also
  derive asteroseismological distances which are in agreement with the
  astrometric measurements of {\it Gaia} for these
  stars.}{Asteroseismological analyses like the one presented in this
  paper could lead to a more complete knowledge of the processes
  occurring during crystallization inside white dwarfs. Also, they
  could make it possible to deduce the core chemical composition of
  ultra-massive white dwarfs, and in this way, to infer their
  evolutionary origin, i.e., if the star has a ONe core and then is
  the result of single-star evolution, or if it harbour a
  carbon/oxygen (CO) core and is the product of a merger of the two
  components of a binary system. However, to achieve these objectives
  it is necessary to find more pulsating ultra-massive WDs, and also
  to carry out additional observations of the already known pulsating
  stars to detect more pulsation periods. Space missions such as {\it
    TESS} can give a great boost in this direction.}

\keywords{stars  ---  pulsations   ---  stars:  interiors  ---  stars: 
          evolution --- stars: white dwarfs}
\authorrunning{C\'orsico et al.}
\titlerunning{Asteroseismology of ultra-massive ZZ Ceti stars}
\maketitle
%_____________________________________________________________________

\section{Introduction}

ZZ Ceti (also called DAV stars) stars are the most  numerous and best studied class of pulsating white dwarf (WD) stars. They are  normal DA WDs with effective temperatures between
$\sim 10\,400$ K and $\sim 12\,400$ K and  logarithm of surface gravities in the range $[7.5 - 9.1]$. These stars exhibit brightness variations  due to nonradial $g$(gravity) modes with low harmonic degree ($\ell\leq 2$)  with periods in the interval $[70-1500]$ s
\citep[][]{2008ARA&A..46..157W,2008PASP..120.1043F,2010A&ARv..18..471A,2019A&ARv..27....7C}.
The first object of this class, R548, was 
discovered to be pulsating by \cite{1968ApJ...153..151L}. From then until now, a large number 
of ZZ Ceti stars have been discovered, initially through specific efforts with observations of 
bright targets  from the ground
\citep{2008PASP..120.1043F}, then from the Sloan Digital 
Sky Survey \citep[SDSS;][]{2000AJ....120.1579Y}, and, in the last years, with the 
{\it Kepler} space  telescope \citep{2016RPPh...79c6901B}, and the {\it Kepler}’s 
second mission K2 \citep{2016PASP..128g5002V}. Currently, there are 260 ZZ Ceti stars known \citep{2019A&ARv..27....7C}, and it is expected that the Transiting Exoplanet Survey Satellite \citep[TESS;][]{2014SPIE.9143E..20R}
will increase this number substantially. 

Asteroseismology is a powerful technique that  offers the exciting prospect of deducing the internal structure of stars by studying their natural frequencies. In the case of pulsating WDs, the first asteroseismic studies of ZZ Ceti stars that compared the observed periods with the theoretical periods computed on a large grid of realistic DA WD models (the so-called {\it forward method}) were carried out by \cite{1998ApJS..116..307B,2001ApJ...552..326B}.  These pioneering works showed that it would be possible, in principle,  to infer the internal chemical structure, the stellar mass, surface gravity, effective temperature, luminosity, radius, seismological distance, and 
rotation rate of ZZ Ceti stars on the basis of the observed pulsation periods. Since then, detailed asteroseismological studies of DAV stars have been carried out, either through the use of fully evolutionary models  \citep{2012MNRAS.420.1462R,2013ApJ...779...58R,2017ApJ...851...60R,2017A&A...599A..21D,2018A&A...613A..46D}, or by using static/parametric models  \citep{2008ApJ...675.1505B,2013MNRAS.429.1585F, 2016MNRAS.461.4059B,2017A&A...598A.109G,2017ApJ...834..136G}. 
Both methods have strengths and weaknesses, but eventually 
they are complementary to each other \citep[see discussion in][]{2019A&ARv..27....7C}.

Of particular interest in this paper is the asteroseismological analysis of 
the rare ultra-massive ZZ Ceti stars ($M_{\star} \gtrsim 1 M_{\sun}$). At variance 
with average-mass ($0.50 \lesssim M_{\star}/ M_{\sun} \lesssim 0.70$) 
and massive ($0.70 \lesssim M_{\star}/ M_{\sun} \lesssim 1.0$) ZZ Ceti stars 
which likely have C/O cores\footnote{There are also the pulsating Extremely 
Low-Mass (ELM) and Low-Mass (LM) WDs, also called ELMVs ($M_{\star} \lesssim 0.30 M_{\sun}$), 
which show H-rich atmospheres and are thought to have cores composed by He.}, ultra-massive ZZ Ceti stars are supposed 
to harbour cores made mostly of O and Ne if they are the result of single-star evolution. However, it cannot be ruled out that ultra-massive WDs
could have CO cores if they are the result of the merger of two WDs \citep{2012ApJ...749...25G}. By virtue of their very 
high masses, these stars are expected to have a large fraction of
their cores crystallized at the effective temperatures characterizing the ZZ Ceti instability strip. The crystallization process is due to Coulomb interactions in very dense plasmas. It was theoretically predicted  
to take place in the cores of WDs  six decades ago \citep{Kirzhnits1960, Abrikosov1961,1961ApJ...134..669S,1968ApJ...151..227V}, 
but it was not until recent  times  that the existence of crystallized WDs was inferred from the study of WD luminosity function of stellar clusters
\citep{2009ApJ...693L...6W,2010Natur.465..194G}, and the galactic field \citep{2019Natur.565..202T}. The effects of crystallization on the pulsational properties of ZZ Ceti 
star models have been studied by \cite{1999ApJ...526..976M,2004ApJ...605L.133M,2004A&A...427..923C,2005A&A...429..277C,2005ApJ...622..572B}. 

In the specific case of ultra-massive WDs with ONe cores, 
the first attempt of studying their pulsational
properties from a theoretical perspective was made by \citet{2004A&A...427..923C}, who 
showed that the forward and 
mean period spacing of ONe-core WDs are markedly different from those 
of CO-core WDs. Recently, \cite{2019A&A...621A.100D} revisited the 
topic by assessing the adiabatic pulsation properties of
ultra-massive DA WDs with ONe cores on the  basis of a new set of fully
evolutionary models generated by \cite{2019A&A...625A..87C}. These 
models  incorporate the most updated physical
ingredients for modelling the progenitor and WD evolution. 
Specifically, the chemical profiles of the WD models of \cite{2019A&A...625A..87C}, which  were  adopted 
from \citet{2010A&A...512A..10S}, are consistent with  
the predictions of the progenitor evolution with stellar masses 
in the range $9.0 < M_{\rm ZAMS}/M_{\sun} < 10.5$ from the ZAMS to the end of the  
the Super Asymptotic Giant Branch (S-AGB) phase.  In addition,  
these models consider, for the first time, the changes in the core 
chemical composition resulting from phase separation due to crystallization, according to 
the predictions of the phase diagram suitable for $^{16}$O and
$^{20}$Ne plasma of \cite{2010PhRvE..81c6107M}. 

\begin{figure}
\includegraphics[clip,width=1.0\columnwidth]{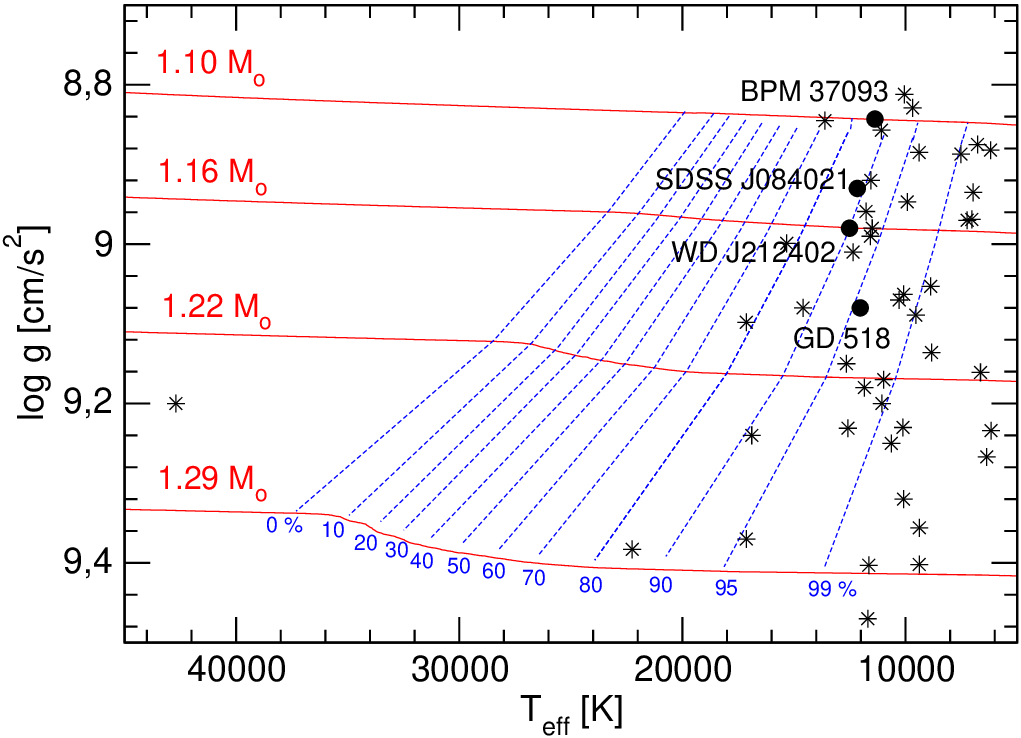}
\caption{Evolutionary tracks (red solid lines) of the ultra-massive DA WD models 
computed by \cite{2019A&A...625A..87C} in the
  $T_{\rm eff}-\log g$ plane. Blue dashed  lines indicate $0, 10, 20,
  30, 40, 50, 60, 70, 80, 90, 95$ and $99 \%$ of crystallized
  mass. The  location of ultra-massive DA WD stars
  \citep{2013ApJS..204....5K,2016MNRAS.455.3413K,2017MNRAS.468..239C}
  are indicated with black star  symbols. The black circles indicate
  the location of the known  ultra-massive ZZ Ceti stars: BPM~37093
  \citep{2016IAUFM..29B.493N}, SDSS J084021 \citep{2017MNRAS.468..239C}, 
  GD 518 \citep{2013ApJ...771L...2H}, and WD J212402 \citep{2019MNRAS.486.4574R}.}
\label{fig-tracks-masivos}
\end{figure}

In this paper, we perform for the first time a detailed asteroseismological analysis of 
the ultra-massive ZZ Ceti stars known up to date on the basis of  the new grid of ONe-core WD models presented in \cite{2019A&A...625A..87C}. At present, there are four objects of this class known: BPM~37093  \citep[$M_{\star}= 1.1~M_\odot$;][]{1992ApJ...390L..89K},  GD~518 \citep[$M_{\star}= 1.24~M_\odot$;][]{2013ApJ...771L...2H}, SDSS~J084021 \citep[$M_{\star}= 1.16~M_\odot$;][]{2017MNRAS.468..239C}, and WD J212402 \citep[$M_{\star}= 1.16 M_{\odot}$;][]{2019MNRAS.486.4574R}. The location of these stars in the spectroscopic Hertzsprung-Russell (HR) diagram is shown in Fig. \ref{fig-tracks-masivos}, along with the evolutionary tracks of \cite{2019A&A...625A..87C}.  The observed pulsation periods of these stars are shown in Tables \ref{table1}, \ref{table3}, \ref{table5}, and \ref{table7}. The star with the richest pulsation spectrum, BPM~37093 (Table \ref{table1}), allows  for a detailed asteroseismological analysis. This star is the main target in the present paper. The remaining stars exhibit just three periods (GD~518 and SDSS~J084021, Tables \ref{table3} and \ref{table5}, respectively) and only one period (WD J212402; Table \ref{table7}). In view of this, for GD~518 and SDSS~J084021 only a preliminary seismological analysis is possible, while for WD~J212402 it is not possible at present to carry out any asteroseismological inference. The stellar models on which we base our study consider time-dependent element diffusion and crystallization with  chemical rehomogeneization due to phase separation. In order to have a set of models suitable for a detailed asteroseismological analysis, we have expanded our set of models by generating new sequences of WD models characterized by H envelopes thinner than the (thick) canonical envelopes. In this way, we  extend the parameter space to be explored  in our asteroseismological analysis. 

The paper is organized as follows. A brief description of the numerical codes  and  the evolutionary models employed is provided in Sect. \ref{numerical}.  In Sect. \ref{pulsation} we present a brief description of the pulsation 
properties of our models. In Section \ref{bpm37093} we perform a detailed asteroseismological analysis of the ultra-massive ZZ Ceti star   BPM~37093, and in Section
\ref{other-massive} we carry out period-to-period fits to the stars GD 518 and SDSS J084021. Finally, in Sect. \ref{conclusions} we summarize the main findings of this work.

\section{Numerical codes and evolutionary models}  
\label{numerical}  

\subsection{Evolutionary and pulsational codes}
\label{evol-pul}

The ultra-massive DA WD evolutionary models employed  in this work were 
computed with the {\tt LPCODE} evolutionary code \cite[see][for detailed
  physical description]{2005A&A...435..631A,2010ApJ...717..897A,
  2010ApJ...717..183R,2016A&A...588A..25M}. This
numerical tool has been employed to study multiple aspects of the
evolution of low-mass stars \citep{2011A&A...533A.139W,
  2013A&A...557A..19A,2015A&A...576A...9A}, the formation of
horizontal branch stars \citep{2008A&A...491..253M}, extremely
low-mass WDs \citep{2013A&A...557A..19A}, AGB and post-AGB evolution \citep{2016A&A...588A..25M}, the evolution of DA WDs \citep{2016ApJ...823..158C} and H-defficient WDs \citep{2017ApJ...839...11C}, among others. 
More recently, the code has been employed to assess the 
impact of the uncertainties in progenitor evolution on 
the pulsation properties and asteroseismological models of ZZ Ceti stars
\citep{2017A&A...599A..21D,2018A&A...613A..46D}. 
The input physics of the version of the {\tt LPCODE} evolutionary code employed 
in this work is described in 
\cite{2019A&A...625A..87C}. We refer the interested reader to that paper 
for details. Of particular importance in this study, is the treatment of crystallization. Theoretical models predict that cool WD stars must crystallize due to the strong
Coulomb interactions in their very dense interiors
\citep{1968ApJ...151..227V}. The two additional energy sources induced by crystallization, 
namely, the release of latent heat, and gravitational energy associated to changes 
in the chemical profiles induced by crystallization, are consistently 
taken into account. The chemical redistribution due to phase separation and the
associated release of energy  have been considered following  \citet{2010ApJ...719..612A}, appropriately modified by \cite{2019A&A...625A..87C} for ONe plasmas.
To assess the enhancement of $^{20}$Ne in the crystallized core, we
used the azeotropic-type phase diagram of
\citet{2010PhRvE..81c6107M}. 

The pulsation code used to compute the nonradial $g$-mode pulsations 
of our complete set of  models is the adiabatic version of the
{\tt LP-PUL} pulsation code described in \citet{2006A&A...454..863C}. 
We did not consider torsional modes, since these modes are characterized by very 
short periods  \citep[up to 20 s; see][]{1999ApJ...526..976M} which 
have never been observed in ZZ Ceti stars. To account for the effects of
crystallization on the pulsation spectrum of $g$ modes, we adopted the
``hard-sphere'' boundary conditions \citep{1999ApJ...526..976M,2005A&A...429..277C}, 
which assume that the amplitude
of the radial displacement of $g$ modes is drastically reduced below the
solid/liquid boundary layer because of the non-shear modulus of the solid,
as compared with the
amplitude in the fluid region
\citep[][]{1999ApJ...526..976M}. The squared Brunt-V\"ais\"al\"a frequency 
($N^2$) for the fluid part of the models is computed as in \cite{1990ApJS...72..335T}. 
The Ledoux term $B$, that explicitly contains the contributions of the chemical 
interfaces to the Brunt-V\"ais\"al\"a frequency, has been appropriately generalized in order
to include the presence of transition regions in which multiple nuclear species vary in abundance.

\begin{figure} 
\includegraphics[clip,width=1.0\columnwidth]{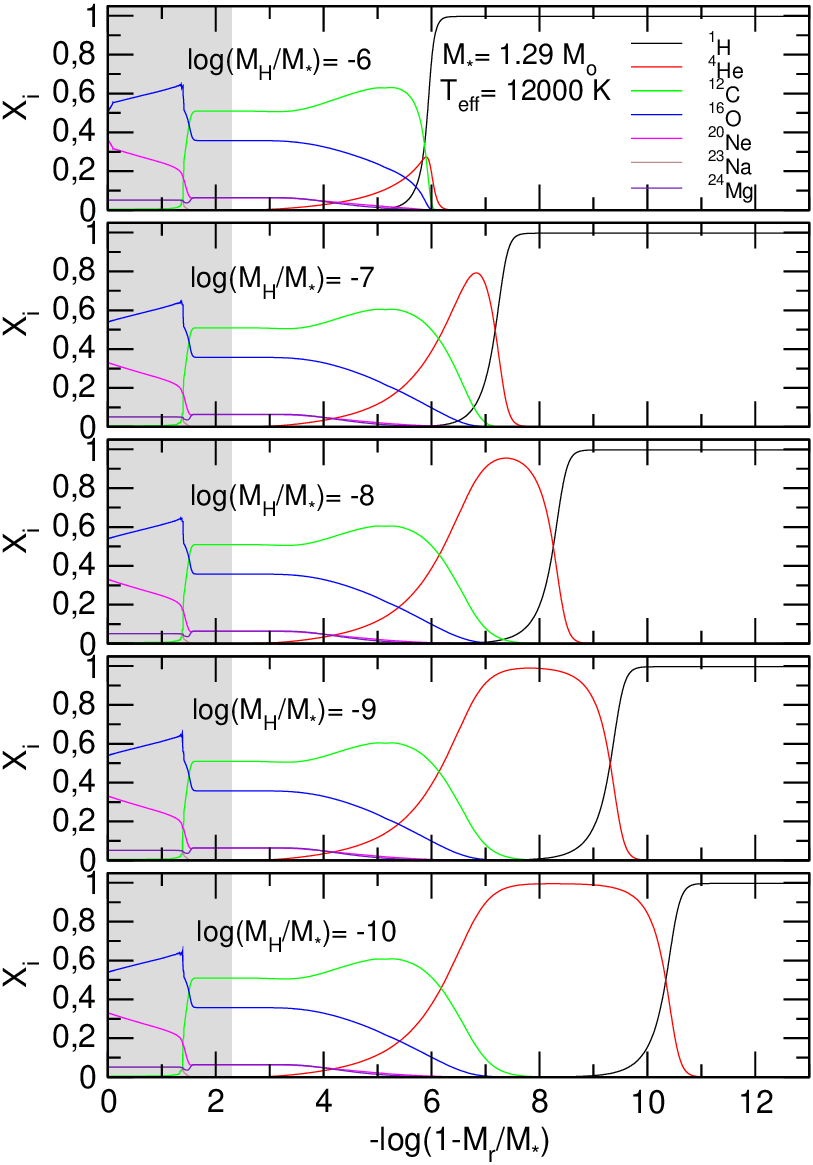} 
\caption{Abundances by mass of $^1$H, $^4$He, $^{12}$C, $^{16}$O, 
$^{20}$Ne, $^{23}$Na, and $^{24}$Mg as  a  function  of  the  fractional  mass, corresponding  to   ONe-core  WD models with $M_{\star}1.29 M_{\sun}$, $T_{\rm eff} \sim 12\,000$ K and
$\log(M_{\rm H}/M_{\star})= -6, -7, -8, -9$, and $-10$ (from top to bottom). The models 
were computed taking into account time-dependent element diffusion, and 
latent heat release and chemical redistribution 
caused by phase separation during crystallization. The solid part of 
the models is emphasized with a gray tone. The crystallized mass fraction 
(in percentage) is $99.5 \%$.}
\label{profiles} 
\end{figure} 

\subsection{The grid of ultra-massive ONe-core WD models}
\label{models}

The asteroseismological analysis presented in this work is based on a set of 
four evolutionary sequences of ultra-massive WD models with stellar masses 
$M_{\star}= 1.10, 1.16, 1.22$, and $1.29~ M_{\sun}$ resulting from the complete evolution of
the progenitor stars through the S-AGB phase \citep{2019A&A...625A..87C}.  
The core  and inter-shell chemical profiles of our models at the start of the  WD
cooling phase were derived from \citet{2010A&A...512A..10S}. The
cores are composed mostly of $^{16}$O and $^{20}$Ne and smaller
amounts of $^{12}$C, $^{23}$Na, and $^{24}$Mg \citep[see Figs. 2 and 3 of ][]{2019A&A...625A..87C}.  Since element
diffusion and gravitational settling operate throughout the WD
evolution, our models develop pure H envelopes.
The He content of our WD sequences is  given by the evolutionary
history of progenitor star, but instead, the H content  of our canonical (thick)
envelopes [$\log(M_{\rm  H}/M_{\star}) \sim -6$] has been set by imposing that the further
evolution  does not lead to H thermonuclear flashes on the WD cooling
track. We have expanded our grid of models by artificially generating new sequences 
harbouring thinner H envelopes [$\log(M_{\rm  H}/M_{\star})= -7, -8, -9, -10$], for each stellar-mass value. This artificial procedure has been done at high-luminosity 
stages of the WD evolution. The resulting transitory effects of this procedure 
become irrelevant much before the models reach the ZZ Ceti regime. 
Details about the 
method to compute the chemical rehomogeneization 
at the core regions during crystallization are given in \cite{2019A&A...625A..87C} and \cite{2019A&A...621A.100D}. The temporal changes of the chemical abundances due to element diffusion are assessed by using a new full-implicit treatment for time-dependent  element  diffusion described in detail in Althaus et al. (2019, submitted). 

In Fig. \ref{profiles} we show the $^1$H, $^4$He, $^{12}$C, $^{16}$O, 
$^{20}$Ne, $^{23}$Na, and $^{24}$Mg chemical
profiles in terms of the fractional mass for $1.29 M_{\sun}$
ONe-core WD models at $T_{\rm eff}\sim 12\,000$ K and H envelope thicknesses 
$\log(M_{\rm H}/M_{\star})= -6, -7, -8, -9$, and $-10$.  Note that a pure He buffer develops as we consider thinner H envelopes (from the top to the bottom panel). 

\begin{figure} 
\includegraphics[clip,width=1.0\columnwidth]{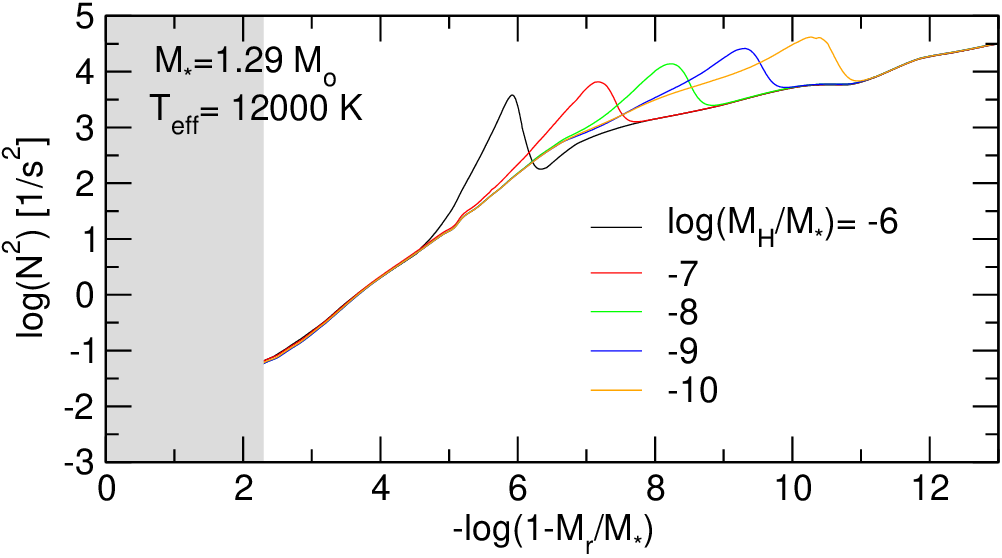} 
\caption{Logarithm of the squared Brunt-V\"ais\"al\"a
  frequency, corresponding to the same ONe-core WD models with
  $M_{\star}= 1.29 M_{\sun}$, $T_{\rm eff} \sim 12\,000$ K and $\log(M_{\rm H}/M_{\star})= -6, -7, -8, -9$, and $-10$ shown in Fig. \ref{profiles}. They gray zone corresponds to the crystallized part of the models.}
\label{bv} 
\end{figure}

At this effective temperature, the chemical rehomogeneization due to crystallization has already finished, 
giving rise to a core where the abundance of $^{16}$O ($^{20}$Ne) increases (decreases) outward. In Fig. \ref{bv} we show the logarithm 
of the squared Brunt-V\"ais\"al\"a frequency corresponding to the same models shown in Fig. \ref{profiles}. The step at the triple chemical transition between $^{12}$C, $^{16}$O, and 
$^{20}$Ne seen in Fig. \ref{profiles} [$-\log(1-M_r/M_{\star}) \sim 1.4$] is within the solid part of the core, thus, it is irrelevant for the mode-trapping properties of these models. This is because, according to the hard-sphere boundary conditions adopted for the pulsations, the eigenfunctions do not penetrate the solid region (gray zone). In view of this, the mode-trapping properties of the models illustrated in Fig. \ref{profiles} and \ref{bv} are entirely determined by the presence  of the He/H transition and the associated bump in the Brunt-V\"ais\"al\"a frequency, which is  located in more external regions for thinner H envelopes (see the next Section).

\section{Pulsation calculations}
\label{pulsation}

\begin{figure*} 
\includegraphics[clip,width=2.0\columnwidth]{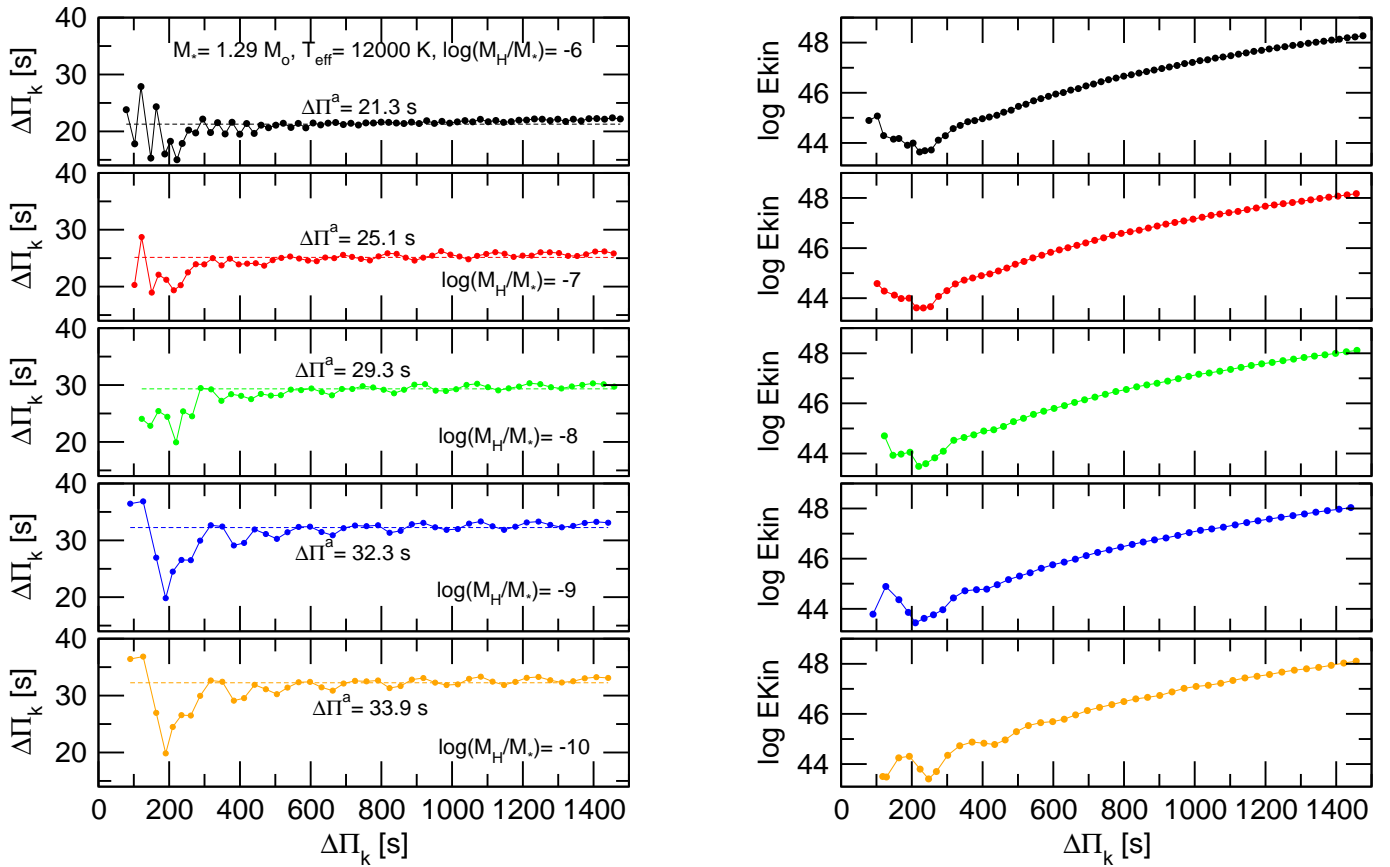}
\caption{Left panels: the forward period spacing, $\Delta \Pi_k$, in
  terms of the pulsation periods, $\Pi_k$, for WD models with
  $M_{\star}= 1.29 M_{\sun}$, $T_{\rm eff} \sim 12\,000$ K and
  different thicknesses of the H envelope.  The thin horizontal dashed lines   correspond to the value of the asymptotic period spacing, $\Delta  \Pi^{\rm a}$.  Right panels: the oscillation kinetic energy versus   the periods for the same WD models shown in the left panel. The normalization $(\delta r/r)_{r= R_{\star}}= 1$ ($\delta r$ being the radial displacement), has been assumed to compute the kinetic energy values.}
\label{DP-EKIN} 
\end{figure*} 

We computed adiabatic pulsation periods of $\ell= 1, 2$ $g$ modes in a range of periods covering the period spectrum that is typically observed in ZZ Ceti stars (70 s $\lesssim \Pi \lesssim 1500$ s).  
We briefly examine the impact of the inclusion of
thin H envelopes on the mode-trapping  properties of our  
ultra-massive WD models. 
Mode trapping of $g$ modes in WDs
is a well-studied mechanical resonance for the mode propagation, 
that acts due to the
presence of density gradients induced by chemical transition regions. 
Specifically, chemical
transition regions, which involve non-negligible jumps in density,
act like reflecting walls that partially trap
certain modes,  forcing  them to oscillate  with larger amplitudes in specific 
regions  and with smaller amplitudes outside those regions 
\citep[see, for details,][]{1992ApJS...80..369B,1992ApJS...81..747B,
1993ApJ...406..661B,2002A&A...387..531C}.
From an observational point of view, a possible signature of mode
trapping in a WD star is the departure from uniform period spacing.
According to the asymptotic theory of stellar pulsations, \emph{in absence of
chemical gradients}, the pulsation periods of $g$ modes  with  high
radial  order  $k$  (long  periods) are expected to be uniformly
spaced with a constant period separation given by \citep{1990ApJS...72..335T}:

\begin{equation} 
\Delta \Pi_{\ell}^{\rm a}= \Pi_0 / \sqrt{\ell(\ell+1)},  
\label{aps}
\end{equation}

\noindent where

\begin{equation}
\Pi_0= 2 \pi^2 \left[\int_{\rm fluid} \frac{N}{r} dr\right]^{-1}.
\label{pcero}
\end{equation}

Actually, the period separation in \emph{chemically stratified} WD models
like the ones considered in this work is not constant, except for very-high 
radial-order modes. We define the forward period spacing as $\Delta
\Pi_k= \Pi_{k+1}-\Pi_k$. The left panels of Fig. \ref{DP-EKIN}
show $\Pi_k - \Delta \Pi_k$ diagrams for the same WD models
depicted in Figs. \ref{profiles} and \ref{bv}. These models
are characterized by $M_{\star}= 1.29 M_{\sun}$ at $T_{\rm eff} \sim 12\,500$ K, and different thicknesses of the H envelope. In each panel, the horizontal
dashed line corresponds to the asymptotic period spacing, computed with Eqs. (\ref{aps}) and (\ref{pcero}).  Models with decreasing H envelope thicknesses
are displayed from top to bottom, starting with the case of the 
canonical envelope.  By examining the plots, several aspects are worth 
mentioning. Firstly, the asymptotic period spacing increases 
for decreasing H envelope thickness. This is because the integral 
in Eq. (\ref{pcero}) for the quantity $\Pi_0$
is smaller for thinner H envelopes, by virtue that the bump in the Brunt-V\"ais\"al\"a
frequency induced by the He/H chemical interface becomes progressively narrow
in the radial coordinate $r$ as this interface is located
at more external layers. Since $\Pi_0$ is larger for
thinner H envelopes, the asymptotic period spacing increases (Eq. \ref{aps}).
$\Delta \Pi_{\ell}^{\rm a}$ experiences an increase between $37\%$ and $60\%$ when
we go from the canonical envelope [$\log(M_{\rm H}/M_{\star})= -6$]
to the thinnest envelope [$\log(M_{\rm H}/M_{\star})= -10$] for this sequence.
Other outstanding feature to be noted from the left panels of
Fig. \ref{DP-EKIN} is connected with the changes in the
mode-trapping properties when we consider H envelopes progressively thinner.
Indeed, we note that for thick envelopes, including the canonical one,
the period-spacing distribution of $g$ modes shows 
a regular pattern of mode trapping with a very short trapping 
cycle ---the $k$ interval between two trapped
modes. When we consider thinner H envelopes, the trapping cycle and the trapping
amplitude increase. A common feature for all the values of $\log(M_{\rm H}/M_{\star})$ considered is that the mode-trapping signatures 
exhibited by $\Delta \Pi_k$ vanish for very large radial orders
(very long periods), in which case $\Delta \Pi_k$
approaches to $\Delta \Pi_{\ell}^{\rm a}$, as predicted by
the asymptotic theory. 

Mode-trapping effects also translate into local maxima and minima in the
kinetic energy of oscillation, $E_{\rm kin}$,  which are usually
associated to  modes that are  partially confined to  the core regions
and  modes that are partially trapped in the envelope. This can be
appreciated in the right panels of Fig. \ref{DP-EKIN}. 
The behaviour described above for $\Delta \Pi_k$
is also found in the case of $E_{\rm kin}$, that is, the mode-trapping
cycle and amplitude increase with decreasing H envelope thickness.

\section{Asteroseismological analysis of BPM~37093}
\label{bpm37093}

\cite{1992ApJ...390L..89K} discovered the first ultra-massive ZZ Ceti star, BPM~37093. This star is characterized 
by  $T_{\rm eff}= 11\,370$ K and $\log g= 8.843$ \citep{2016IAUFM..29B.493N}.
Detailed theoretical computations carried out by \cite{1997ApJ...487L.191W},  \cite{1998PhDT........21M}, and \cite{1999ApJ...526..976M}, suggested that 
BPM~37093 should have a crystallized core. This star has been the target of two multisite observing
campaigns of the Whole Earth Telescope \citep[WET;][]{1990ApJ...361..309N}. Preliminary results from these campaigns were published
by \cite{2000BaltA...9...87K}. The 1998 observations (XCov 16)
revealed a set of regularly spaced pulsation frequencies in the
range 1500-2000 $\mu$Hz. The 1999 observations (XCov 17) revealed a
total of four independent modes, including two new modes and
two that had been seen in the previous campaign. By comparing pulsation amplitudes in the UV to the optical spectra, \cite{2000PhDT.......103N} identified the harmonic degree of the  
BPM~37093 pulsation modes, concluding that they can not be $\ell = 3$ and that most of the modes must be $\ell = 2$. \cite{2004ApJ...605L.133M} obtained
new single-site observations of BPM~37093 from the Magellan
6.5 m telescope on three nights in 2003 February. These data
showed evidence of five independent modes, all of which had
been detected in the two previous multisite campaigns.
\cite{2005A&A...432..219K} reported on WET observations of BPM~37093 obtained in 1998 and 1999 and,
on the basis of a simple analysis of the average period spacing, they 
concluded that a large fraction of the total stellar mass of the star 
should be crystallized. On the basis of asteroseismological techniques,  
\cite{2004ApJ...605L.133M} reported to have "measured" the  
crystallized mass fraction in BPM~37093 and determined a value $\sim 90$ \%. 
However, employing similar asteroseismological methods, \cite{2005ApJ...622..572B} questioned those conclusions, suggesting instead that the percentage of crystallized mass of BPM~37093 probably should be between 32 \% and 82 \%. In the next sections, we carry out a detailed astroseismological analysis that involves the assessment of a mean period spacing and its comparison with the theoretical values, and also  
period-to-period fits with the intention of finding an asteroseismological model.

\begin{figure}
	\centering
	\includegraphics[width=1.0\columnwidth]{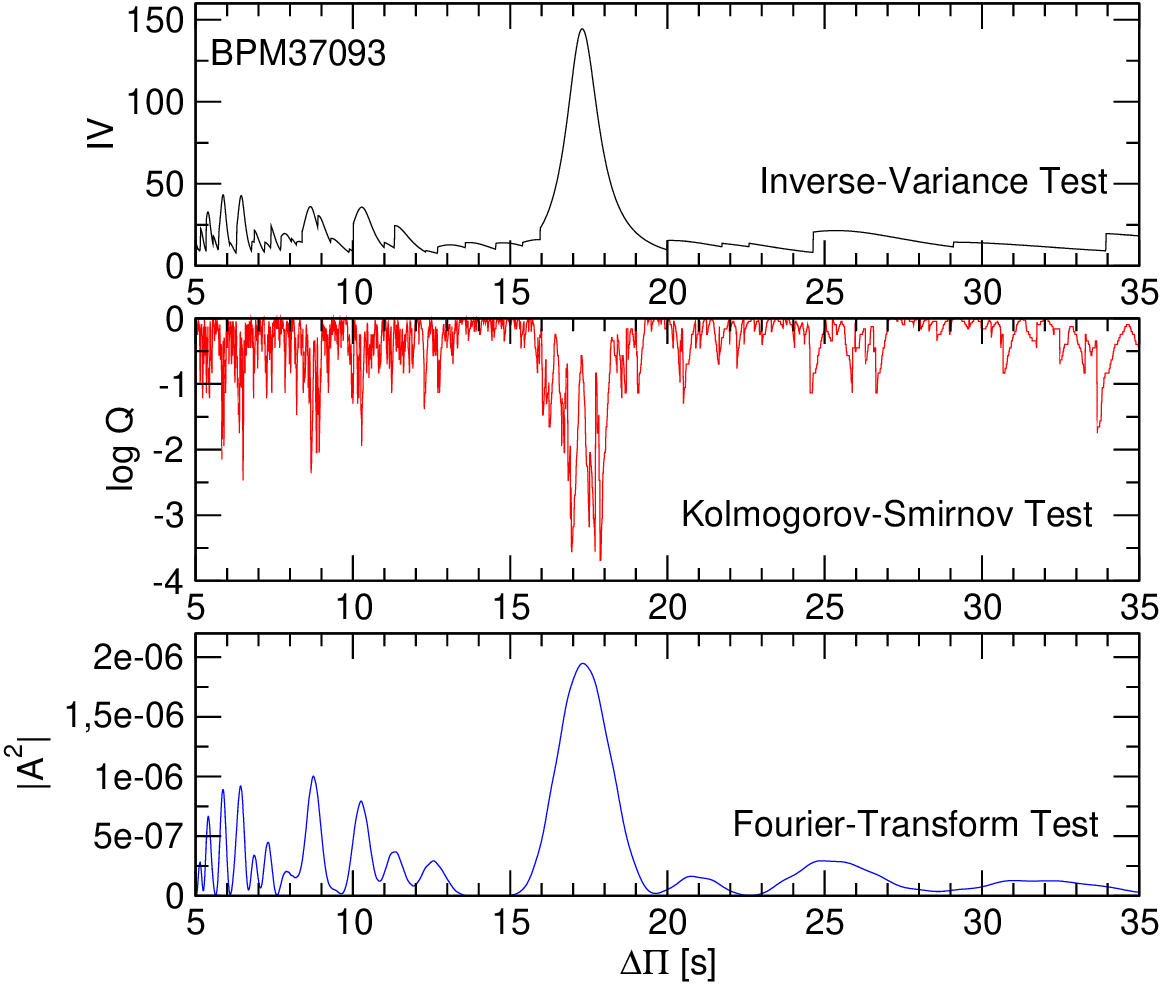}
	\caption{I-V  (upper panel),  K-S (middle panel),  and  F-T (bottom panel) significance  tests   applied   to   the   period   spectrum  of BPM~37093 to search for a constant period spacing. The periods used here are the 8 periods shown in Table \ref{table1}.}	\label{tests-bpm37093}
\end{figure}

\subsection{Period spacing}
\label{ps-bpm37093}

For the asteroseismological analysis of this star, we adopt the set of eight modes considered by \cite{2004ApJ...605L.133M} (see Table \ref{table1}). This list of periods is based on the set of periods detected by \cite{2000PhDT.......103N}. We  searched  for  a  constant  period  spacing  in  the  data of BPM~37093 by  using  the  Kolmogorov-Smirnov  
\citep[K-S; see][]{1988IAUS..123..329K}, the inverse variance \citep[I-V; see][]{1994MNRAS.270..222O} and the 
Fourier Transform \citep[F-T; see][]{1997MNRAS.286..303H} significance tests. 
In the K-S test, any uniform or at least systematically non-random period spacing in the period 
spectrum of the star will appear as a minimum in $Q$. In the I-V test, a maximum of the inverse variance will indicate a constant period spacing. Finally, in the F-T test, we calculate the Fourier transform of a Dirac comb function (created from the set of observed periods), and then we plot the square of the amplitude of the resulting function in terms of the inverse of the frequency. And once again, a maximum in the square of the amplitude will indicate a constant period spacing. In Fig. \ref{tests-bpm37093}  we show the results of applying the tests to the set of periods of Table \ref{table1}.  The three tests indicate the existence of a mean period spacing of about $17$ s.
According to our set of models, the asymptotic period spacing (Eq. \ref{aps}) for ultra-massive DA WDs with masses between $1.10$ and $1.29 M_{\sun}$ and effective temperatures within the ZZ Ceti instability strip ($13\,500 {\rm K}- 10\,500$ K) 
varies between $\sim 22$ s 
and $\sim 34$ s for $\ell= 1$, and between $\sim 12$ s 
and $\sim 19$ s for $\ell= 2$. Clearly, the period spacing evidenced by the 3 tests for BPM~37093 corresponds to modes 
$\ell = 2$. This indicates that the period spectrum of this star is dominated by quadrupole modes, being this in concordance with the finding of \cite{2000PhDT.......103N}. 
By averaging the period spacing derived from the three statistical tests, we found 
$\Delta \Pi_{\ell= 2}= 17.3 \pm 0.9$ s. 

\cite{2016IAUFM..29B.493N} expanded the set of periods of \cite{2000PhDT.......103N} for BPM~37093 by employing  Gemini  South time-series combined  with  simultaneous  time-series photometry  from  Mt.  John  (New  Zealand),  SAAO,  PROMPT,
and Complejo Astronómico El Leoncito (CASLEO, Argentina), providing a list of 13 periods (see their Table 1). Averaging two pairs of periods, the list gives 11 periods, of which 8 are the same as in \cite{2004ApJ...605L.133M}, and the remaining 3 periods are new periods with values 624.2 s, 641.0 s, 660.8 s. We have applied the three statistical tests to this expanded list of periods, but we have not found a clear period spacing. This is because the 3 new 
periods do not fit well into the regular pattern of periods with $\ell = 2$ shown
in Table \ref{table1}.

\begin{table}[]
\centering
\caption{The independent frequencies and periods in the data of BPM~37093 from \cite{2004ApJ...605L.133M}, along with the theoretical periods, harmonic degrees, radial orders, and period differences  of the best-fit model described in Sect. \ref{ptp-bpm37093}.}
\begin{tabular}{cc|cccc}
\hline
\hline
$\Pi^{\rm O}$  & $\nu$       & $\Pi^{\rm T}$  & $\ell$ & $k$ & $\delta_i$\\ 
$[$sec$]$ & [$\mu$Hz] &  [sec]         &        &     &  [sec] \\ 
\hline
511.7 & 1954.1 & 512.4 & 2 & 29 & $-0.7$\\
531.1 & 1882.9 & 531.9 & 1 & 17 & $-0.8$\\
548.4 & 1823.5 & 548.1 & 2 & 31 & $0.3$\\ 
564.1 & 1772.7 & 565.3 & 2 & 32 & $-1.2$\\
582.0 & 1718.2 & 583.0 & 2 & 33 & $-1.0$\\
600.7 & 1664.9 & 599.9 & 2 & 34 & $0.8$\\
613.5 & 1629.9 & 613.8 & 1 & 20 & $-0.3$\\
635.1 & 1574.6 & 632.2 & 2 & 36 & $2.9$\\
\hline
\end{tabular}
\label{table1}
\end{table}

\subsection{Average of the computed period spacings}
\label{averaged}

In principle, the stellar mass of   pulsating WDs can be derived by comparing the average of the period spacings (or the asymptotic period spacing\footnote{Generally, the use of the asymptotic period spacing (computed according to Eq. \ref{aps}),  instead of the average of the computed period spacings, can lead to an overestimation of the stellar mass, except for stars that pulsate with very high radial orders, such as PNNV stars \citep{2008A&A...478..175A}.}) computed from a grid of models with different masses, effective temperatures, and  envelope thicknesses with the mean period spacing exhibited by the star, if present. This method takes full advantage of the fact that the period spacing of DBV (pulsating DB WDs) 
and GW Vir stars (pulsating PG1159 stars) primarily 
depends on the stellar mass and the effective temperature, and very weakly on the thickness of the He envelope in the case of DBVs \citep[see, e.g.,][]{1990ApJS...72..335T} and the thickness 
of the C/O/He envelope in the case of the GW Vir stars \citep{1994ApJ...427..415K}. In the case of ZZ Ceti stars, however, the the average of the period spacings and the asymptotic period spacing depend on the stellar mass, the effective temperature, and the thickness of the H envelope with a comparable sensitivity. Consequently, the method is not --in principle-- directly applicable to ZZ Ceti stars due to the intrinsic degeneracy of the dependence of $\Delta \Pi$ with the three parameters $M_{\star}, T_{\rm eff},$ and $M_{\rm H}$ \citep{2008PASP..120.1043F}.

\begin{figure*}
	\centering
	\includegraphics[width=1.8\columnwidth]{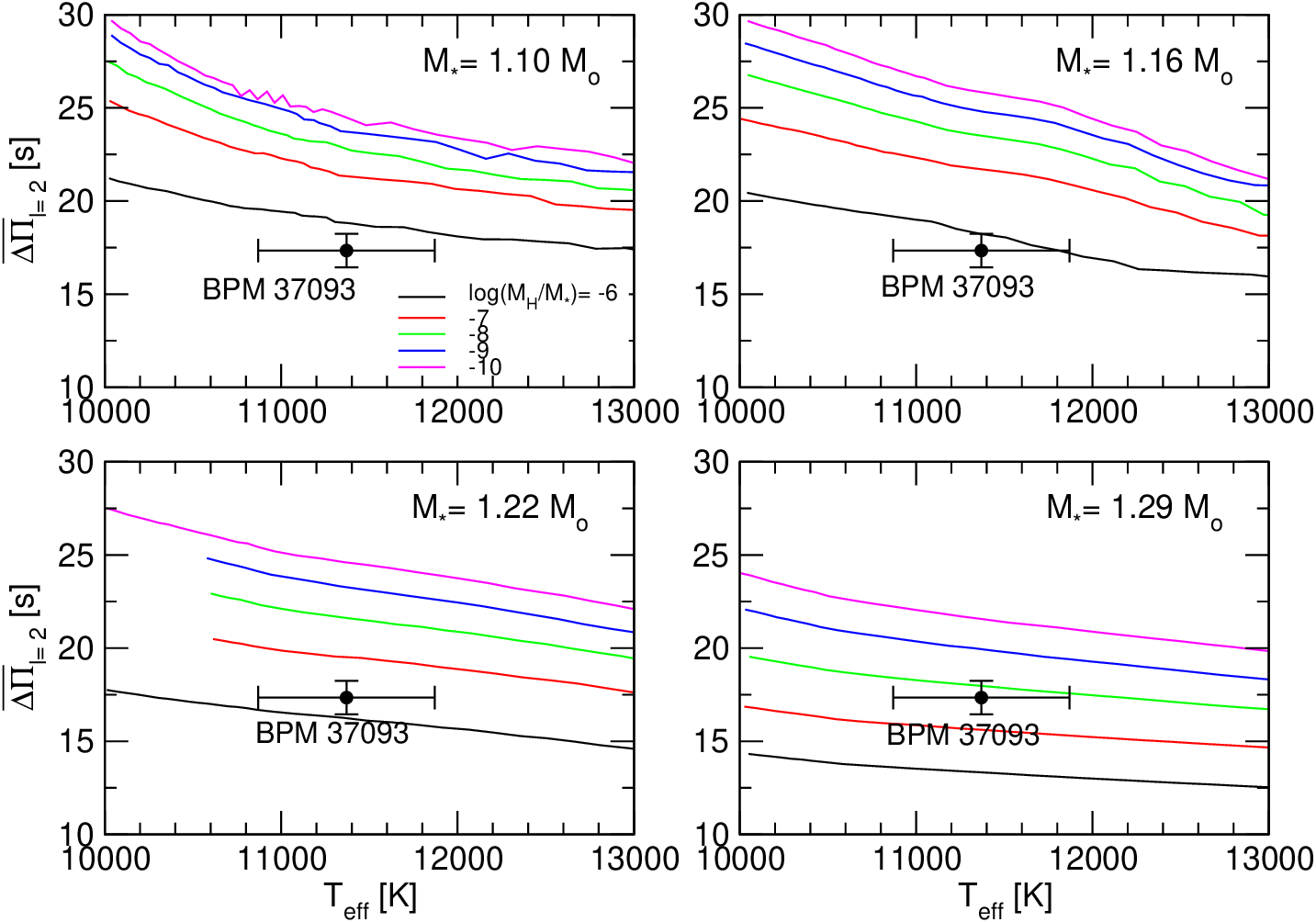}
	\caption{Comparison between the quadrupole ($\ell= 2$) period spacing derived for BPM~37093 ($\Delta \Pi= 17.3 \pm 0.9$ s), and the average of the computed $\ell= 2$ period spacings, $\overline{\Delta \Pi}_{\ell= 2}$, for all the considered stellar masses and different H-envelope thicknesses, in terms of the effective temperature.}	
	\label{aps-bpm37093}
\end{figure*}

In spite of this caveat, we tried to derive the stellar mass of BPM~37093 from the measured quadrupole period spacing. To this end, we assessed the average quadrupole period spacings computed for our models
as $\overline{\Delta  \Pi}_{\ell=2}= (n-1)^{-1}  \sum_{k}^n  \Delta \Pi_k  $,
where $\Delta \Pi_k$ is the forward period spacing 
for $\ell= 2$ modes
and $n$ is the number of theoretical periods considered from the model.
For BPM~37093, the observed periods are in the range [511,635] s. 
In computing the averaged period spacings for the models, however, we have considered the range $[500,1400]$\,s, that is, 
we adopted a longer upper limit of this range of periods in order 
to better sample the period spacing of modes within the asymptotic regime.  In Fig.~\ref{aps-bpm37093} we  show the  run of  the average of  the computed period  spacings  ($\ell= 2$) in  terms of  the effective temperature for our ultra-massive DA  WD evolutionary sequences for all the thicknesses of the H envelope, along with the observed period spacing for BPM~37093.  As can be appreciated from the figure, it is not possible in this instance to put very strong constraints on the mass of BPM~37093, and the only thing that can be assured is that the mass of the star could be $M_{\star}= 1.16 M_{\sun}$ with a thick (canonical) H envelope [$\log(M_{\rm H}/M_{\star})= -6$], but it could also be as massive as $M_{\star}= 1.29 M_{\sun}$ and with a  H envelope 100 times thinner [$\log(M_{\rm H}/M_{\star})= -8$]. This degeneracy of the solutions could be eliminated with the help of period-to-period fits. In the next section, we address this issue.

\subsection{Period-to-period fits} 
\label{ptp-bpm37093}

Here, we search for a pulsation model that best matches the \emph{individual} pulsation periods of BPM~37093. The  goodness of  the  match  between the   theoretical pulsation  periods
($\Pi_k^{\rm  T}$) and  the observed   individual  periods ($\Pi_i^{\rm  O}$) is measured by means  of a merit function defined
as: 

\begin{equation}
\label{chi}
\chi^2(M_{\star}, M_{\rm   H}, T_{\rm   eff})=   \frac{1}{N} \sum_{i=1}^{N}   \min[(\Pi_i^{\rm   O}-   \Pi_k^{\rm  T})^2], 
\end{equation}

\noindent where $N$ is the number of observed periods. The WD model that shows the lowest value of $\chi^2$, if exists, is adopted as the ``best-fit model''.  We assess the function
$\chi^2=\chi^2(M_{\star}, M_{\rm H}, T_{\rm eff})$ for stellar masses of $1.10$, $1.16$, $1.22$, and $1.29\ M_{\sun}$.
For the effective temperature we cover a range of
$15000 \gtrsim T_{\rm eff} \gtrsim 10000\ $ K. Finally, 
for the  H-envelope thickness we adopt the values $\log(M_{\rm H}/M_{\star})= -6, -7, -8, -9, -10$. The  quality of  our period fits is assessed by means of the average of the absolute period differences, $\overline{\delta}= \left(\sum_{i= 1}^N |\delta_i| \right)/N$,  where $\delta_i= \Pi_i^{\rm   O}-\Pi_k^{\rm  T}$, and by the root-mean-square residual, $\sigma= \sqrt{(\sum_{i= 1}^N  |\delta_i|^2)/N}= \sqrt{\chi^2}$.

\begin{figure}
	\centering
	\includegraphics[width=1.0\columnwidth]{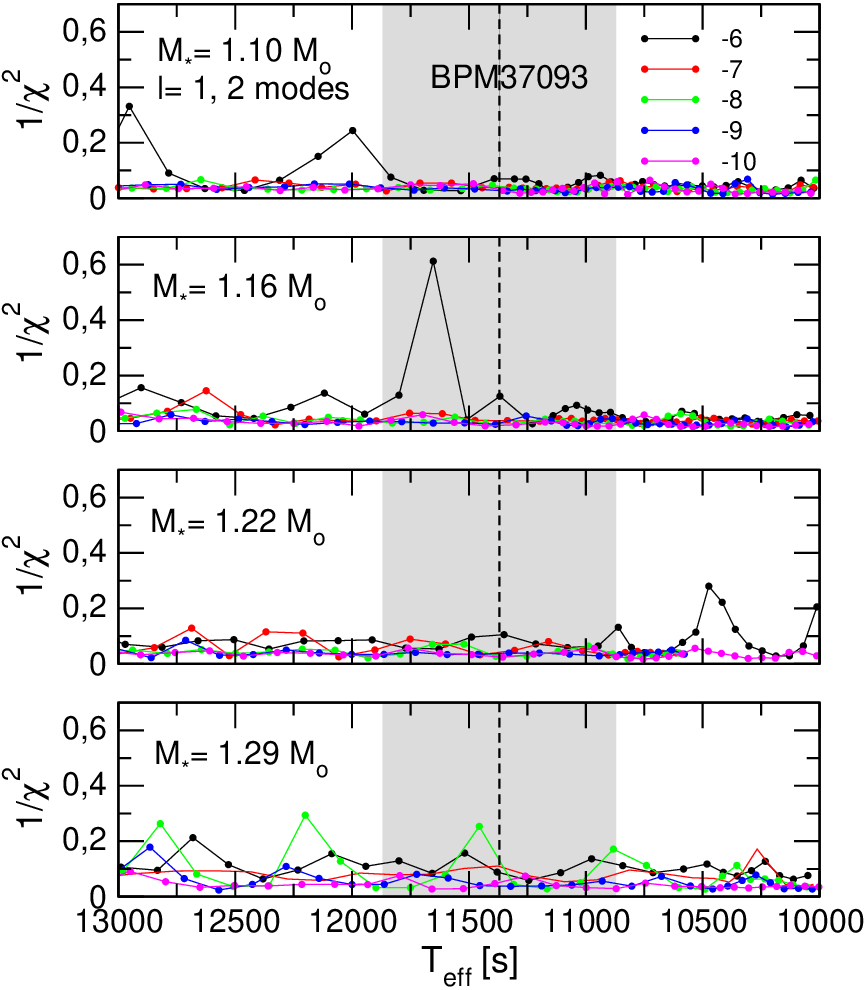}
	\caption{The inverse of the quality function of the 
	period fit in the case in which we allow the periods to be associated to $\ell= 1$ and $\ell= 2$ modes in terms of the effective temperature for the ultra-massive DA WD model sequences with different stellar masses ($M_{\star}$) and H envelope thicknesses [$\log(M_{\rm H}/M_{\star})$], as indicated. The vertical dashed line and the gray strip correspond to the spectroscopic effective temperature of BPM~37093 and its uncertainties ($T_{\rm eff}= 11\,370 \pm 500$ K). Note the strong maximum in $(\chi^2)^{-1}$ for $M_{\star}= 1.16 M_{\sun}$ and $\log(M_{\rm H}/M_{\star})= -6$  at $T_{\rm eff} \sim 11\,650$ K. This corresponds to our "best-fit" model (see text for details).}	
	\label{chi2-l1l2-bpm37093}
\end{figure}

\begin{figure}
	\centering
	\includegraphics[width=1.0\columnwidth]{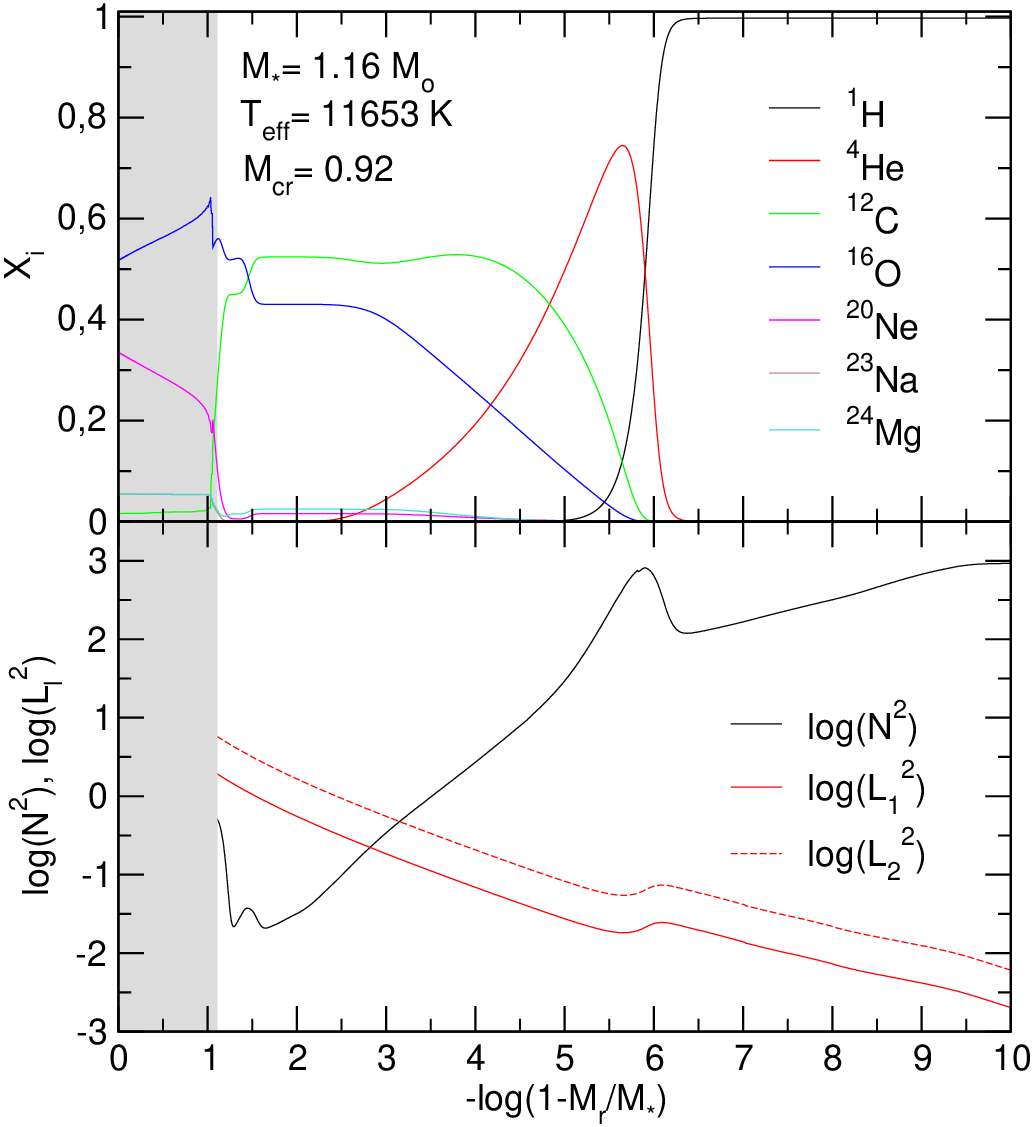}
	\caption{The internal chemical structure (upper panel), and the
 squared Brunt-Va\"is\"al\"a and Lamb frequencies for $\ell= 1$ and 
 $\ell= 2$
 (lower panel) corresponding to our best-fit ultra-massive DA WD model for BPM~37093
 with a stellar mass $M_*= 1.16\,M_{\odot}$, an effective temperature
 $T_{\rm eff}= 11\,653$\,K, a H envelope mass of $\log(M_{\rm H}/M_*)  \sim -6$, and a crystallized mass fraction of $M_{\rm cr}= 0.92 M_{\star}$.}	
	\label{XBV-best-fit}
\end{figure}

We assumed two possibilities for the mode identification: {\it (i)} that  all of  the observed periods correspond to $g$ modes associated to $\ell= 1$, and {\it (ii)} that  the observed periods correspond to a mix of $g$ modes
associated to $\ell= 1$ and $\ell= 2$. We first 
considered the 8 periods employed by \cite{2004ApJ...605L.133M} (see Table \ref{table1}). The case {\it (i)} did not show clear solutions compatible with BPM~37093 in relation to its spectroscopically-derived effective temperature. Instead, the case {\it (ii)}, in which we allow the periods of the star to be associated to a combination of $\ell= 1$ and $\ell= 2$ modes, resulted in a clear seismological solution
for a WD model with $M_{\star}= 1.16 M_{\sun}$, $T_{\rm eff}= 11\,650$ K and $\log(M_{\rm H}/M_{\star})= -6$, 
as it can be appreciated from Fig. \ref{chi2-l1l2-bpm37093}. 
In Table \ref{table1} we show the periods of the best-fit model along with the harmonic degree, the radial order, and the period differences. For this model, we obtain $\overline{\delta}= 1.00$ s and
$\sigma= 1.28$ s. In order to have an indicator of the quality of the period fit, we computed the Bayes Information Criterion \citep[BIC;][]{2000MNRAS.311..636K}:
\begin{equation}
{\rm BIC}= N_{\rm p} \left(\frac{\log N}{N} \right) + \log \sigma^2,
\end{equation}
\noindent where $N_{\rm p}$ is the number of free parameters of the models, and $N$ is the number of observed periods. The smaller the value of BIC, the
better the quality of the fit. In our case, $N_{\rm p}= 3$ (stellar
mass, effective temperature, and thickness of the H envelope), 
$N= 8$, and $\sigma= 1.28$\,s. 
We obtain ${\rm BIC}= 0.55$, which means that our fit is very good.
In Table \ref{table2}, we list the main characteristics of the best-fit model. The seismological stellar mass is in good agreement with the 
spectroscopic inference based on the evolutionary tracks of \cite{2019A&A...625A..87C}. The quadrupole ($\ell= 2$) mean period spacing of our 
best fit model is $\overline{\Delta \Pi}= 17.63$ s, in excellent agreement with the mean period spacing derived for BPM~37093 
($\Delta \Pi= 17.3 \pm 0.9$ s). 
 Fig. \ref{XBV-best-fit} depicts the chemical profiles (upper panel) and the propagation diagram (lower panel) corresponding to the best-fit model of BPM~37093. Our best-fit model has $\sim 92 \%$ of its mass 
in crystalline state.

We repeated the process of period fit considering the preliminary set of 13 periods observed by \cite{2016IAUFM..29B.493N}, but we did not find a clear seismological solution neither when we considered the case {\it (i)} nor when we adopted the case {\it (ii)}. 

\subsection{Internal uncertainties}

We   have   assessed   the  uncertainties   in   the   stellar   mass
($\sigma_{M_{\star}}$),  the thickness  of the  H  envelope ($\sigma_{M_{\rm H}}$),  and the effective  temperature ($\sigma_{T_{\rm  eff}}$), of
the  best-fit model by  employing the  expression \citep{1986ApJ...305..740Z,2008MNRAS.385..430C}:

\begin{equation}
\sigma_i^2= \frac{d_i^2}{(S-S_{\rm 0})},
\end{equation}

\noindent where $S_{\rm 0}\equiv  \chi^2(M_{\star}^{\rm 0}, M_{\rm H}^{\rm 0},
T_{\rm eff}^{\rm  0})$ is  the minimum of  $\chi^2$ which is  reached at
$(M_{\star}^{\rm 0}, M_{\rm H}^{\rm  0},T_{\rm eff}^{\rm 0} )$ corresponding
to the best-fit  model, and $S$ is the value of  $\chi^2$ when we change
the parameter $i$  (in this case, $M_{\star}, M_{\rm  H}$, or $T_{\rm eff}$)
by an amount $d_i$, keeping  fixed the other parameters.  The quantity
$d_i$  can  be evaluated  as  the  minimum step  in  the  grid of  the
parameter $i$.   We obtain the following uncertainties:
$\sigma_{M_{\star}}\sim  0.014  M_{\sun}$,  $\sigma_{M_{\rm  H}}  \sim  6.3 \times 10^{-7} M_{\star}$,  and  $\sigma_{T_{\rm eff}}\sim  40$  K.   The uncertainty in $L_{\star}$ 
is derived from the width of the maximum in the function ($1/\chi^2$) 
in terms of $L_{\star}$. We obtain 
$\sigma_{L_{\star}} \sim 5.8 \times 10^{-4} L_{\sun}$. The
uncertainties in  $R_{\star}$ and $g$ are derived
from the uncertainties in $M_{\star}$, $T_{\rm eff}$, and $L_{\star}$. 

\begin{table}
\centering
\caption{The main characteristics of BPM~37093. The second column  
corresponds to spectroscopic and astrometric results, whereas the third  
column presents results from the asteroseismological model of this work.} 
\begin{tabular}{l|cc}
\hline
\hline
Quantity                     & Spectroscopy                     & Asteroseismology             \\
\hline
$T_{\rm eff}$ [K]            & $11\,370\pm 500^{\rm (a)}$       & $11\,650\pm 40$   \\
$M_{\star}/M_{\sun}$         & $1.098\pm 0.1^{\rm (b)}$                  & $1.16 \pm 0.014$    \\ 
$\log g$ [cm/s$^2$]          & $8.843 \pm 0.05^{\rm (a)}$        & $8.970 \pm 0.025 $  \\ 
$\log (L_{\star}/L_{\sun})$  & ---                              & $-3.25\pm 0.01$ \\  
$\log(R_{\star}/R_{\sun})$   & ---                              & $-2.234\pm 0.006$ \\  
$\log(M_{\rm H}/M_{\star})$  & ---                              & $-6 \pm 0.26$   \\  
$\log(M_{\rm He}/M_{\star})$  & ---                             & $-3.8$ \\           
$M_{\rm cr}/M_{\star}$    &   $0.935^{\rm (b)}$                              & $0.923$ \\  
$X_{^{16}{\rm O}}$ cent.               & ---                               &  $0.52$ \\     
$X_{^{20}{\rm Ne}}$ cent.               &  ---                            & $0.34$  \\
\hline
\hline
Quantity                 &  Measured                & Asteroseismology \\  
\hline
%$\Delta \Pi^{\rm a}_{\ell=1}$ [s]     & ---                            & 29.92 \\    
%$\Delta \Pi^{\rm a}_{\ell=2}$ [s]     & ---                            & 17.28 \\    
$\overline{\Delta \Pi}_{\ell= 1}$ [s]  &---                             & 29.70  \\    
$\overline{\Delta \Pi}_{\ell= 2}$ [s]  &  $17.3\pm0.9$                 & 17.63 \\    
%BC [mag]                     & $-5.81_{-0.21}^{+0.23}$          & $-5.89_{-0.04}^{+0.08}$      \\ 
%$M_{\rm V}$ [mag]            & $7.55_{-0.51}^{+0.74}$           & $7.79_{-0.10}^{+0.03}$       \\
%$M_{\rm bol}$ [mag]          & $1.74$                           & $1.9_{-0.14}^{+0.11}$        \\
%$A_{\rm  V}$ [mag]           & $0.19$                           & $0.071$ \\ 
\hline
\hline
Quantity                     & Astrometry (\emph{Gaia})                    & Asteroseismology             \\
\hline
$d$  [pc]                     & $14.81\pm0.01$                            &     $11.38\pm0.06$       \\ 
$\pi$ [mas]                   & $67.52\pm0.04$                           &        $87.87\pm0.40$       \\ 
\hline
\hline
\end{tabular}
\label{table2}
{\footnotesize  References: (a)  \cite{2016IAUFM..29B.493N}. (b) \cite{2019A&A...625A..87C}}
\end{table}

Table \ref{table2} includes the parameters of the best-fit model along with the  uncertainties derived above. These are formal uncertainties related to the process of searching for the asteroseismological model, and therefore they can be considered as \emph{internal} uncertainties inherent to the asteroseismological process. 

\subsection{Asteroseismological distance}
\label{ast-dist}

We employ the effective temperature and gravity of our best-fit model 
to infer the absolute $G$ magnitude ($M_{\rm G}$) of BPM~37093 in the \emph{Gaia} photometry (D. Koester, private communication). We find $M_{\rm G}= 13.53$ mag.
On the other hand, we obtain the apparent magnitude $m_{\rm G}= 13.8$ mag from the \emph{Gaia} Archive\footnote{({\tt https://gea.esac.esa.int/archive/}).}. According to the well-known expression  
$\log d=  (m_{\rm G} -M_{\rm G}+5)/5$,  we obtain  $d= 11.38\pm0.06$ pc and a parallax $\pi= 87.87\pm0.40$ mas. These asteroseismological distance and parallax are somewhat different as compared with those provided directly by \emph{Gaia}, that is, $d= 14.81\pm0.01$ pc and $\pi= 67.52\pm0.04$ mas. 
However, we note that the uncertainties in the asteroseismological distance and parallax come mainly from the uncertainties in the effective temperature and the logarithm of the gravity of the best-fit model ($\sim 40$ K and $\sim 0.025$), which are admittedly small because they are just internal errors. Realistic estimates of these errors are probably much higher (see above Section). That said, we believe that with more realistic estimates of the uncertainties in $T_{\rm eff}$ and $\log g$, and thus in the errors in the asteroseismological distance and parallax, the agreement with the astrometric values could substantially improve.

\subsection{Rotation period}

If the stellar rotation is slow and rigid, the rotation frequency $\Omega$ of the WD is connected with the 
frequency  splitting $\delta \nu$ through the coefficients $C_{k, \ell}$ ---that depend on the details of the stellar structure--- and the values of $m$ ($-\ell, \cdots, -1, 0, +1, \cdots, +\ell$), by means of the expression $\delta \nu = m(1-C_{k,\ell})\ \Omega$ \citep{1989nos..book.....U}. The period at 564.1 s in Table \ref{table1} is actually the average of two very close observed periods which are assumed to be the components $m= -1$ (562.6 s) and $m= +1$ (565.5 s) of a  rotationally split $\ell= 2$ mode \citep[see][]{2016IAUFM..29B.493N}. Here, it is assumed that the remainder components of the quintuplet ($m = -2, 0, +2$) 
are not visible for some unknown reason. Under this hypothesis, we derive a frequency 
splitting of $\delta \nu= 4.55 \mu$Hz. Making the same assumption for the 
pair of observed periods at 633.5 s and 636.7 s \citep[see][]{2016IAUFM..29B.493N}, which, averaged, give the period 635.1 s in Table \ref{table1}), we have $\delta \nu= 3.95 \mu$Hz.  
For our best-fit model for BPM~37093, we find that the 564~s and 635~s modes have $C_{k,\ell=2}\sim 0.166$. Using this value for $C_{k,\ell}$, and the averaged frequency  splitting, $\overline{\delta \nu}= 4.25 \mu$Hz, we obtain a rotation period of $\sim 55$\,h. 
This rotation period is consistent with  the  rotation-period values inferred from asteroseismology for WD stars \citep[see Table 10 of][]{2019A&ARv..27....7C}.
We also can estimate what the rotation period would be if these periods were the components $m= -1$ and $m= +1$ of $\ell= 1$ modes instead of $\ell= 2$ modes. In that case, we would have $C_{k,\ell=2}\sim 0.498$ from the best-fit model, and then the rotation period should be of $\sim 33$\,h. 

\subsection{Comparison with previous analyses}
\label{prev-BPM}

\cite{2004ApJ...605L.133M} carried out a parametric asteroseismological analysis on BPM~37093 
on the basis of the eight periods listed in the first column of Table \ref{table1}.
These authors employed DA WD models characterized by chemical transition regions resulting from the assumption of diffusive equilibrium. The free parameters of the analysis are the crystallized mass fraction (that is, the location of the inner boundary conditions for the pulsations, which coincides with the liquid/solid interface), the He and H envelope thickness, and the effective temperature. The authors consider pure C- and O-core WDs, and three fixed stellar-mass values.  
They obtain a family of asteroseismological models characterized by different stellar parameters, but all of them with 90 \% of the mass crystallized. A second parametric asteroseismological analysis of BPM~37093 was performed independently by \cite{2005ApJ...622..572B}, who employed DA WD models with some improved aspects; for example, updated opacities, chemical transitions resulting from time-dependent element diffusion, and cores made of CO in addition to pure C and O cores. In addition, the models of \cite{2005ApJ...622..572B} do not consider the crystallized mass fraction as a free parameter, but instead, the value is fixed for each model and results from the predictions of the EoS. The results of this analysis largely differ from those of \cite{2004ApJ...605L.133M}. Indeed, \cite{2005ApJ...622..572B} found a set of optimal asteroseismological models characterized by a percentage of crystallized mass in the range 32-82 \%. These authors emphasize that the information contained in the eight periods employed in both analyses is not enough to unravel the core chemical structure nor to derive the percentage of crystallized mass of this star, due to the fact that the modes are characterized by high radial orders and therefore, they are in the asymptotic regime of $g$-mode pulsations. 
The strong differences of the results of the works by \cite{2004ApJ...605L.133M}
and \cite{2005ApJ...622..572B} could be due to the fact that
\cite{2004ApJ...605L.133M}  only varied the crystallized mass fraction in increments of 10 \% (i.e., 10\%, 20\%, 30\%, $\cdots$, 80\%, 90\%). Using a finer 
grid in the increments of the crystallized mass fraction could result in many 
other possible best-fit solutions, potentially more in agreement with the larger set of solutions found by \cite{2005ApJ...622..572B}.

The DA WD models employed in the present paper are substantially different as compared with those employed by \cite{2004ApJ...605L.133M} and \cite{2005ApJ...622..572B}, particularly regarding the core chemical structure and composition. In fact, while those authors consider cores made of pure C, pure O, and mixtures of 50 \% of C and 50 \% of O, in the present analysis we consider cores made of O and Ne with evolving chemical structures as predicted by fully evolutionary computations. In addition, our asteroseismological approach, which is based on fully evolutionary models, largely differs from that adopted in the mentioned works, that is, the employment of structure models with a number of adjustable free parameters to search for the optimal asteroseismological models.  
For these reasons, a direct comparison of our results with those of \cite{2004ApJ...605L.133M} and \cite{2005ApJ...622..572B} is not possible. However, 
we can emphasize that our analysis favours a WD model with a large fraction of mass in solid phase ($\sim 92 \%$), more in line with the results of  \cite{2004ApJ...605L.133M}. Also, the identification of the harmonic degree $\ell$ and the radial order $k$ of the pulsation modes for the asteroseismological solutions are similar. Indeed, our analysis predicts that most of the modes are quadrupole modes, except the modes with periods at 531.1 s and 613.5 s, which are dipole modes. In the case of  
\cite{2004ApJ...605L.133M}, most of the the modes are $\ell= 2$, except modes with periods 582.0 s and 613.5 s which are $\ell= 1$ modes. Finally, \cite{2005ApJ...622..572B} predict that most of the modes are $\ell= 2$, except the mode with periods 613.5 s, which is a $\ell= 1$ mode.  Regarding the radial order of the modes, in our case we obtain $29 \leq k \leq 36$, whereas both \cite{2004ApJ...605L.133M} and \cite{2005ApJ...622..572B} analyses predict $28 \leq k \leq 35$.    The surprising agreement of the identification of the radial order $k$ of the modes according to \cite{2004ApJ...605L.133M} and \cite{2005ApJ...622..572B} as compared with the current analysis (differing only by 1) could be due to the fact that 
our best-fit model for BPM~37093 has a large fraction of mass crystallized, 
so that $g$-mode pulsations are insensitive to the ONe-core chemical 
features,  and thus, the pulsational properties of the model resemble those of a model with a similar mass but with a CO core.

\section{Other ultra-massive ZZ Ceti stars}  
\label{other-massive}

There are three other pulsating ultra-massive ZZ Ceti stars known to date, apart from BPM~37093. They are GD~518, SDSS~J084021.23+522217.4, and J212402.03$-$600100.0. In contrast to BPM~37093, these three stars show only a few pulsation periods (see Tables \ref{table3}, \ref{table5}, and \ref{table7}), which prevent us from finding a period spacing for these star. Also, the scarcity of periods inhibits us from carrying out a detailed asteroseismological analysis as in the case of BPM~37097. Then, we will limit ourselves to perform a preliminary analysis of period-to-period fits for GD 518 and SDSS~J084021.23+522217.4. The star J212402.03$-$600100.0  is excluded from this analysis because it only has a single detected period.

  \begin{table}[]
\centering
\caption{The independent frequencies in the data of GD~518 from \cite{2013ApJ...771L...2H} along with the theoretical periods, harmonic degrees, radial orders, and period differences of the best-fit model.}
\begin{tabular}{cc|cccc}
\hline
\hline
$\Pi^{\rm O}$ [sec] & $\nu$ [$\mu$Hz] & $\Pi^{\rm T}$ [sec]&$\ell$&$k$&$\delta_i$ [sec]\\ 
\hline
$440.2\pm1.5$ & $2271.7\pm7.6$ &439.55&2& 29 &0.70\\
$513.2\pm2.4$ & $1948.6\pm9.2$ &514.10&2& 34 &$-0.90$\\
$583.7\pm1.5$ & $1713.3\pm4.5$ &583.09&1& 22 &0.61\\
\hline
\end{tabular}
\label{table3}
\end{table}  

\begin{table}
\centering
\caption{Same as Table \ref{table2}, but for GD~518.} 
\begin{tabular}{l|cc}
\hline
\hline
Quantity                     & Spectroscopy                     & Asteroseismology             \\
\hline
$T_{\rm eff}$ [K]            & $12\,030\pm 210^{\rm (a)}$       & $12\,060\pm 38$   \\
$M_{\star}/M_{\sun}$         & $1.198^{\rm (b)}$         & $1.22 \pm 0.03$    \\ 
$\log g$ [cm/s$^2$]          & $9.08 \pm 0.06^{\rm (a)}$        & $9.15 \pm 0.021 $  \\ 
$\log (L_{\star}/L_{\sun})$  & ---                              & $-3.34\pm 0.01$ \\  
$\log(R_{\star}/R_{\sun})$   & ---                              & $-2.31\pm 0.008$ \\  
$\log(M_{\rm H}/M_{\star})$  & ---                              & $-6 \pm 0.24$   \\  $\log(M_{\rm He}/M_{\star})$  & ---                             & $-4$ \\           
$M_{\rm cr}/M_{\star}$    & $0.955^{\rm (b)}$                              & $0.971$ \\  
$X_{^{16}{\rm O}}$ cent.               & ---                               &  $0.53$ \\     
$X_{^{20}{\rm Ne}}$ cent.               &  ---                            & $0.32$  \\
\hline
\hline
%Quantity                 &  Measured                & Asteroseismology \\  
%\hline
%$\Delta \Pi^{\rm a}_{\ell=1}$ [s]     & ---                            & 29.92 \\    
%$\Delta \Pi^{\rm a}_{\ell=2}$ [s]     & ---                            & 17.28 \\    
%$\overline{\Delta \Pi}_{\ell= 1}$ [s]  &---                             & 29.70  \\    
%$\overline{\Delta \Pi}_{\ell= 2}$ [s]  &  $17.3\pm0.9$                 & 17.63 \\    
%BC [mag]                     & $-5.81_{-0.21}^{+0.23}$          & $-5.89_{-0.04}^{+0.08}$      \\ 
%$M_{\rm V}$ [mag]            & $7.55_{-0.51}^{+0.74}$           & $7.79_{-0.10}^{+0.03}$       \\
%$M_{\rm bol}$ [mag]          & $1.74$                           & $1.9_{-0.14}^{+0.11}$        \\
%$A_{\rm  V}$ [mag]           & $0.19$                           & $0.071$ \\ 
%\hline
%\hline
Quantity                     & Astrometry (\emph{Gaia})                    & Asteroseismology             \\
\hline
$d$  [pc]                     & $64.57\pm0.3$                            &      $49.80\pm0.06$       \\ 
$\pi$ [mas]                   & $15.48\pm0.08$                           &       $20.08\pm0.03$       \\ 
\hline
\hline
\end{tabular}
\label{table4}
%\label{table:GD518}
{\footnotesize  References: (a)  \cite{2013ApJ...771L...2H}; (b) \cite{2019A&A...625A..87C}}
\end{table}

\begin{figure}
	\centering
	\includegraphics[width=1.0\columnwidth]{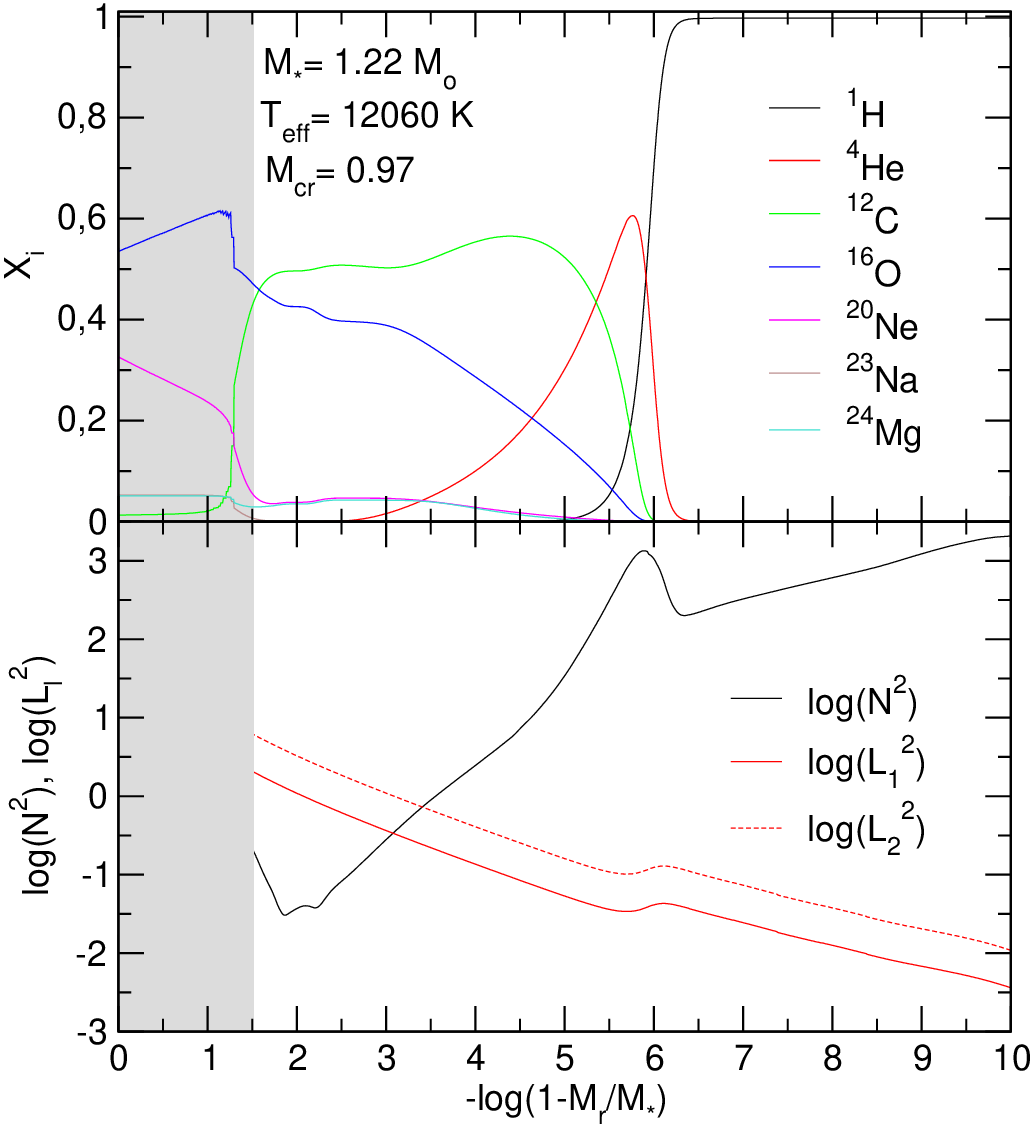}
	\caption{Same as in Fig. \ref{XBV-best-fit}, but for the asteroseismological 
	best-fit model of GD~518.}	
	\label{XBV-best-fit-GD518}
\end{figure}

\subsection{GD~518}

Pulsations in WD J165915.11+661033.3 (GD~518) were first detected by \citet{2013ApJ...771L...2H}. Model-atmosphere fits to this star indicate that it 
is located in the ZZ Ceti instability strip with $T_{\rm eff} \sim 12\,030$ K and $\log g \sim 9.08$, which would correspond to a mass of $1.20 M_{\odot}$   if the ONe-core WD models from \citet{2005A&A...441..689A} are used, or $1.23 M_{\odot}$ if the CO-core WD models from \citet{1995LNP...443...41W} are employed. 
The value of the stellar mass of the star is $M_{\star}= 1.198 M_{\sun}$ if the evolutionary tracks of 
ONe-core WD models of \cite{2019A&A...625A..87C} are adopted. To date, no asteroseismological analysis has been performed to this star. Our period-to-period fits
for this star indicate that our best fit model ---the one which minimizes the merit function from Eq. (\ref{chi})--- is characterized by a value of 
$\chi^2=  0.56$, $\overline{\delta}= 0.74$ s, $\sigma= 0.75$ s, and BIC$= 0.22$, and has a stellar mass of $1.22 M_{\odot}$ and $T_{\rm eff}= 12\,060$ K (see Table \ref{table4} and Fig. \ref{XBV-best-fit-GD518}). The stellar mass of the asteroseismological model is consistent with the spectroscopic mass derived from the evolutionary tracks of \cite{2019A&A...625A..87C}.

 The asteroseismological distance and parallax inferred for GD~518, derived in the same way than for BPM~37093, are $d= 49.80\pm0.06$ pc and $\pi= 20.08\pm0.03$ mas. These values are somewhat different than  those provided by $Gaia$, that is, $d= 64.57\pm0.3$ pc and $\pi=15.48\pm0.08$ mas. The agreement between
these sets of values could improve if we could employ more realistic values for the uncertainties in $T_{\rm eff}$ and $\log g$ of the asteroseismological model for GD~518, in a similar way than for BPM~37093 (see discussion at the end of Sect. \ref{ast-dist}).

\subsection{SDSS J084021.23+522217.4}

This ultra-massive ZZ Ceti star was discovered by \citet{2017MNRAS.468..239C} from a sample of DA WD from the SDSS DR7 and DR10. Model-atmosphere fits indicate $T_{\rm eff} \sim 12 \,160$ K, log $g \sim 8.93$ and $M_{\star} \sim 1.16 M_{\odot}$. These results are in good agreement with the preliminary asteroseismological analysis performed by the same authors, where their best-fit CO-core WD model has $M_{\star}= 1.14 M_{\sun}$, $M_{\rm H}= 5.8\times 10^{-7}M_{\star}$, $M_{\rm He}= 4.5\times 10^{-4}M_{\star}$, $0.50 \leq M_{\rm cr}/M_{\star} \leq 0.70$ and $11\, 850 \leq T_{\rm eff} \leq 12\, 350$ K. 

Our best fit model is characterized by  $\chi ^2= 0.14$, $\overline{\delta}= 0.37$ s, $\sigma= 0.38$ s, and BIC$=-0.37$ with one period (797.4 s) identified as a $\ell= 2$ mode, and the remaining periods identified as $\ell= 1$ modes. The derivation of the stellar parameters gives $T_{\rm eff}=12\,550$ K, $M_{\star}= 1.10 M_{\odot}$, $M_{\rm H}/M_{\star}= 1.02\times 10^{-7}$, $M_{\rm He}/M_{\star}= 3.0\times 10^{-4}$, $ M_{\rm cr}/M_{\star}= 0.81$, with a central $^{20}$Ne abundance of $0.52$. The stellar mass 
derived from the asteroseismological model is somewhat smaller than the value 
derived spectroscopically on the basis of the evolutionary tracks of \cite{2019A&A...625A..87C}. On the other hand, the disagreement regarding the mass of the crystallized part of the core as compared with the result found by \citet{2017MNRAS.468..239C}
is because here we are employing ONe-core WD models, whereas those authors consider CO-core WD models. When searching for the best-fit model with all periods assumed to be  associated to $\ell= 1$ modes, we found the same best-fit model as in the previous analysis, but with a poorer quality function ($\chi ^2=1.56$).

The asteroseismological distance and parallax inferred for this star are $d= 89.95\pm0.08$ pc and $\pi= 11.12\pm0.01$ mas, which differ from the $Gaia$ values, $d= 138.50\pm4.0$ pc and $\pi= 7.22\pm0.21$ mas.
Again, a better estimate of the uncertainties of the effective temperature and gravity of the asteroseismological model could contribute to bring the asteroseismological distance and parallax values closer to those derived by {\it Gaia}.

\begin{table}[]
\centering
\caption{The independent frequencies in the data of SDSS~J084021.23+522217.4 from \cite{2017MNRAS.468..239C} along with the theoretical periods, harmonic degrees, radial orders, and period differences of the best-fit model.}
\begin{tabular}{cc|cccc}
\hline
\hline
$\Pi^{\rm O}$ [sec] & $\nu$ [$\mu$Hz] & $\Pi^{\rm T}$ [sec]&$\ell$&$k$&$\delta_i$ [sec]\\ 
\hline
$172.7\pm0.4$ & $5790.4$ & 172.23&1 & 3 & 0.47 \\
$326.6\pm1.3$ & $3061.8$ & 326.88&1 & 8  & $-0.28$\\
$797.4\pm8.0$ & $1254.14$ & 797.76&2 & 40 & $-0.36$\\
\hline
\end{tabular}
\label{table5}
\end{table}

\begin{table}
\centering
\caption{Same as Table \ref{table2}, but for SDSS J084021.23+522217.4.} 
\begin{tabular}{l|cc}
\hline
\hline
Quantity                     & Spectroscopy                     & Asteroseismology             \\
\hline
$T_{\rm eff}$ [K]            & $12\,160\pm 320^{\rm (a)}$       & $12\,550\pm 70.$   \\
$M_{\star}/M_{\sun}$         & $1.139^{\rm (b)}$          & $1.10 \pm 0.04$    \\ 
$\log g$ [cm/s$^2$]          & $8.93 \pm 0.07^{\rm (a)}$        & $8.84 \pm 0.02$  \\ 
$\log (L_{\star}/L_{\sun})$  & ---                              & $-3.02\pm 0.01$ \\  
$\log(R_{\star}/R_{\sun})$   & ---                              & $-2.18\pm 0.005$ \\  
$\log(M_{\rm H}/M_{\star})$  & ---                              & $-7 \pm 0.21$   \\  $\log(M_{\rm He}/M_{\star})$  & ---                             & $-3.5$ \\           
$M_{\rm cr}/M_{\star}$    & $0.945^{\rm (b)}$                                 & $0.813$ \\  
$X_{^{16}{\rm O}}$ cent.               & ---                    & $0.52$ \\     
$X_{^{20}{\rm Ne}}$ cent.               &  ---                  & $0.31$  \\
\hline
\hline
\hline
Quantity                     & Astrometry (\emph{Gaia})                    & Asteroseismology             \\
\hline
$d$  [pc]                     & $138.50\pm4.0$                            &     $89.95\pm0.08$      \\ 
$\pi$ [mas]                   & $7.22\pm0.21$                           &        $11.12\pm0.01$       \\ 
\hline
\hline
\end{tabular}
\label{table6}
{\footnotesize  References: (a)  \cite{2017MNRAS.468..239C}; (b) \cite{2019A&A...625A..87C}}
\end{table}

\begin{figure}
	\centering
	\includegraphics[width=1.0\columnwidth]{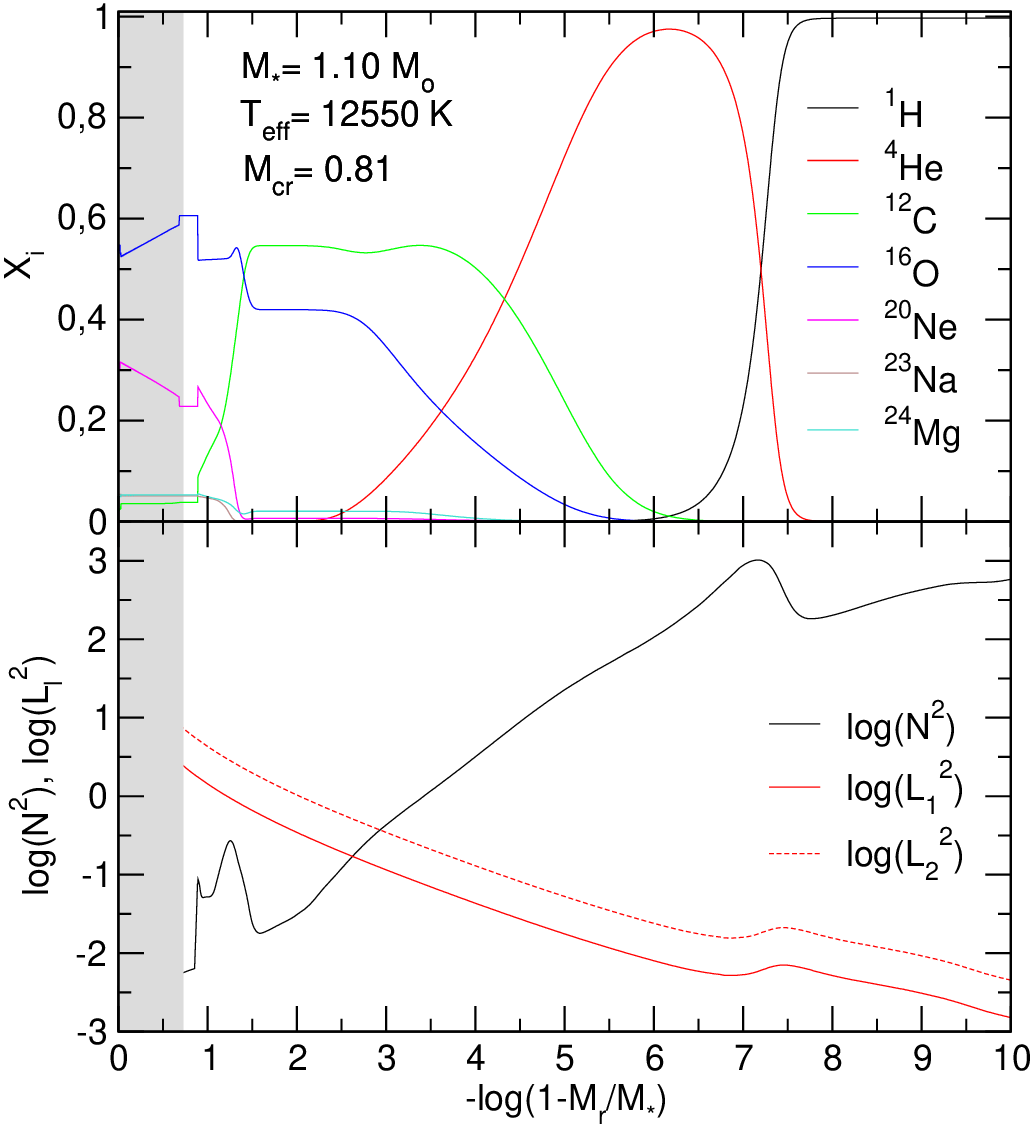}
	\caption{Same as in Fig. \ref{XBV-best-fit}, but for the asteroseismological 
	best-fit model of SDSS J084021.23+522217.4.}	
	\label{XBV-best-fit-SDSSJ084021}
\end{figure}

 \begin{table}[]
\centering
\caption{The single frequency in the data of WD J212402.03$-$600100.0 from \cite{2019MNRAS.486.4574R}.}
\begin{tabular}{cc}
\hline
\hline
$\Pi^{\rm O}$ [sec] & $f$ [$\mu$Hz]  \\ 
\hline
357   & 2801           \\ 
\hline
\end{tabular}
\label{table7}
\end{table}  

\subsection{WD~J212402}

The variability of WD~J212402 was discovered by \cite{2019MNRAS.486.4574R} from time-series GALEX space-telescope observations. This star has $T_{\rm eff}= 12\,510$ K and $\log g= 8.98$ \citep{2019MNRAS.482.4570G}. The stellar mass
of the star is $M_{\star}= 1.16 M_{\sun}$ and the crystallized mass fraction should be of $M_{\rm cr}/M_{\star} \sim 0.90$ according to the evolutionary tracks of \cite{2019A&A...625A..87C}. Unfortunately, only a single period  
has been detected (Table \ref{table7}), preventing us from attempting an asteroseismological analysis. It would be very important to have additional observations of this star to detect more pulsation periods.

\section{Summary and conclusions}
\label{conclusions}

In this paper, we have conducted for the first time an asteroseismological 
study of the ultra-massive ZZ Ceti stars known hitherto 
by employing  an expanded set of grid of ONe-core WD models presented in \cite{2019A&A...625A..87C}. The stellar models on which this study is based consider crystallization with  chemical rehomogeneization due to phase separation. We have included ultra-massive WD models
with different thicknesses of the H envelope, with the aim of 
expanding the parameter space in our asteroseismological exploration.

For the ultra-massive ZZ Ceti star BPM~37093, we have carried out a detailed
asteroseismological analysis that includes the derivation of a mean period spacing of $\sim 17$ s, which is associated to $\ell= 2$ $g$ modes.  We have not been able, however, to infer the stellar mass of the star by comparing the 
observed period spacing with the averaged theoretical period spacings. This is due to  the intrinsic degeneracy of the dependence of $\Delta \Pi$ with the three parameters $M_{\star}, T_{\rm eff}$ and $M_{\rm H}$. On the other hand, 
we have derived a best-fit model for the star, by considering their individual pulsation periods. This model is characterized by $T_{\rm eff}= 11\,650$ K, $M_{\star}= 1.16 M_{\sun}$, $\log(M_{\rm H}/M_{\star})= -6$, and $M_{\rm cr}/M_{\star}= 0.92$ (see Table \ref{table2}).  In addition, we have derived an asteroseismological distance of 11.38 pc, which somewhat differs from the astrometric distance   
measured by {\it Gaia}, of 14.81 pc. Finally, a rotation period of 55 h has been inferred, under the assumption that the modes that exhibit frequency splittings are associated to $\ell= 2$ modes. 
For the ultra-massive ZZ Ceti stars GD~518 and SDSS~J084021, which exhibit only 
three periods, we have performed period-to-period fits, and we find asteroseismological models whose characteristics are listed in Tables \ref{table4} and \ref{table6}. In particular,
this analysis predict that the crystallized mass fraction of these stars 
are $M_{\rm cr}/M_{\star}= 0.97$ (GD~518) and $M_{\rm cr}/M_{\star}= 0.81$ (SDSS~J084021).  The asteroseismological distances inferred for these stars (50 pc and 90 pc, respectively) are 
somewhat different to the distances measured by {\it Gaia} (65 pc and 139 pc, respectively). Finally, for the ultra-massive ZZ ceti star WD~J212402, which exhibits one single period, it is not possible to do any kind of asteroseismological inference at this stage.

 Tables \ref{table2}, \ref{table4}, and \ref{table6} include the parameters of the best-fit models for BPM~37093, GD~518, and SDSS J084021.23+522217.4, respectively. We note that for two of the three stars studied, that is BPM~37093 and GD~518, the percentage of crystallization is larger than 90\% by mass, which is larger than the mass of the ONe core. Thus, since $g$-mode pulsations only sample the non-crystallized regions, and since these regions are dominated by O, C, He and H, it is not surprising that the best-fit seismological models are consistent with prior studies which assumed CO cores, particularly in the case of BPM~37093.

In Tables \ref{table2}, \ref{table4}, and \ref{table6} we have included 
the formal uncertainties related to the process of searching for the asteroseismological model, and therefore they can be considered as \emph{internal} uncertainties inherent to the asteroseismological process.  An estimation of more realistic uncertainties in the structural quantities that characterize the asteroseismological models of these stars ($T_{\rm eff}, M_{\star}, M_{\rm H}, 
M_{\rm He}, R_{\star}$, etc) is very difficult to obtain, since they depend on the uncertainties affecting the physical processes of the progenitor evolution. 
An estimate of the impact of the uncertainties in the prior evolution on the structural parameters of the asteroseismological models has been carried out by \cite{2017A&A...599A..21D,2018A&A...613A..46D} for ZZ Ceti stars of 
intermediate masses  harbouring CO cores. These authors derive typical 
uncertainties of $\Delta M_{\star}/M_{\star} \lesssim 0.05$, $\Delta T_{\rm eff} \lesssim 300$ K and a factor of two in the thickness of the H envelope. While we can not directly extrapolate these results to our analysis of ultra-massive DA WD models with ONe cores, we can adopt them  as representative of the real uncertainties affecting the parameters of our asteroseismological models for BPM~37093, GD~518, and SDSS J084021.23+522217.4.

In this paper, we have assumed that ultra-massive WDs ($M_{\star} \gtrsim 1 M_{\sun}$) come from single-star evolution and must have ONe cores. However, it cannot be discarded that these objects are the result of mergers of 
two WDs \citep[the  so-called  "double  degenerate scenario"; see, e.g.,][]{2012ApJ...749...25G,2012MNRAS.427..190S} in a binary system, in which case it is expected that they have CO cores. The study of the evolutionary and pulsational properties of ultra-massive WDs resulting from WD+WD mergers is beyond the scope of the present paper and will be the focus of a future investigation.

We close the article by emphasizing the need of new photometric observations from the ground or from space (e.g., TESS) in order to find more variable ultra-massive WDs, and also to re-observe the already known objects (for instance WD~J212402) in order to find more periods. This will result in reliable asteroseismological analyses that could yield valuable information about the crystallization processes in WDs. Also, it could be possible to derive the core chemical composition and, in turn, to infer their
evolutionary origin ---that is, either single-star  evolution or binary-star evolution with the merger of two WDs.

\begin{acknowledgements}
 We  wish  to  acknowledge  the  suggestions  and
comments of an anonymous referee that strongly improved the original
version of this work. We gratefully acknowledge Prof. Detlev Koester 
for providing us with a tabulation of the absolute magnitude of DA WD 
models in the \emph{Gaia} photometry.  Part of this work was 
supported by AGENCIA through the Programa de Modernizaci\'on Tecnol\'ogica BID 1728/OC-AR, and by the PIP
  112-200801-00940 grant from CONICET.  This research has made use of
  NASA's Astrophysics Data System.
\end{acknowledgements}

\bibliographystyle{aa}
\bibliography{paper_bibliografia.bib}
%\begin{thebibliography}{}
%\bibitem[Auri\`ere(1982)]{aur82} Auri\`ere, M.  1982, \aap,
%    109, 301
%\end{thebibliography}

\end{document}